\documentclass[fleqn,11pt,a4paper]{article}

\usepackage{epsfig,amsmath,amssymb,bm}
\usepackage{color}
\def\w#1{\,\hbox{#1}} % for extra space and different font on units
\newcommand{\bfr}{{\mathbf r}}
\newcommand{\bfC}{{\mathbf C}}
\newcommand{\bft}{{\mathbf t}}
\newcommand{\bfn}{{\mathbf n}}
\newcommand{\bfb}{{\mathbf b}}
\newcommand{\bfT}{{\mathbf T}}
\newcommand{\bfN}{{\mathbf N}}
\newcommand{\bfB}{{\mathbf B}}

\newcommand{\bfOm}{{\boldsymbol{ \Omega}}}
\newcommand{\cross}{{\mathbf \times}}
\newcommand{\bfi}{{\mathbf i}}
\newcommand{\bfj}{{\mathbf j}}
\newcommand{\bfk}{{\mathbf k}}

\newcommand{\bfx}{{\mathbf x}}
\newcommand{\bfy}{{\mathbf y}}
\newcommand{\bvs}{{\underline{s}}}
\newcommand{\bvx}{{\underline{x}}}
\newcommand{\bvkappa}{{\underline{\kappa}}}
\newcommand{\bvtau}{{\underline{\tau}}}
\newcommand{\bvrho}{{\underline{\rho}}}
\newcommand{\bvl}{{\underline{l}}}

\newcommand{\Om}{{\Omega}}
\newcommand{\Mcal}[1]{\ensuremath{\mathcal{#1}}}

\newcommand{\Lie}[1]{\ensuremath{\mathcal{L}_{#1}}}
\newcommand{\Mdual}[1]{\ensuremath{\tilde{#1}}}

\newcommand{\Hodge}{\ensuremath{\star}}
\newcommand{\ParHodge}{\ensuremath{\#}}
\newcommand{\PerpHodge}{\ensuremath{\#_\perp}}
\newcommand{\PD}{\ensuremath{\partial}}
\newcommand{\Set}[1]{\{#1\}}
\newcommand{\Kkappa}{\Theta}
\newcommand{\Intder}[1]{\iota_{#1}}
\newcommand{\Tensor}{\ensuremath{\otimes}}
\newcommand{\Bdry}{\PD}

\newcommand{\Bcon}{\quad\forall\, p\in\Mcal{B}}
\newcommand{\Invg}{g^{-1}}
\newcommand{\Hlambda}{\hat{\lambda}}
\newcommand{\Hs}{\hat{s}}
\newcommand{\Hx}{\hat{x}}
\newcommand{\Hrho}{\hat{\rho}}
\newcommand{\Hkappa}{\hat{\kappa}}
\newcommand{\Htau}{\hat{\tau}}
\newcommand{\bvHs}{\hat{\underline{s}}}
\newcommand{\bvHx}{\hat{\underline{x}}}
\newcommand{\bvHrho}{\hat{\underline{\rho}}}
\newcommand{\bvHkappa}{\hat{\underline{\kappa}}}
\newcommand{\bvHtau}{\hat{\underline{\tau}}}

\newcommand{\BesselJ}[1]{J_{#1}}
\newcommand{\BR}{{\mathbf R}}
\newcommand{\BZ}{{\mathbf Z}}
\newcommand{\BD}{{\mathbf D}}
\newcommand{\bfgamma}{\bm{\gamma}}
\newcommand{\bfGamma}{\bm{\Gamma}}

\newcommand{\Tstarpar}{(T^*_p\Mcal{M})^\parallel}
\newcommand{\Tstarperp}{(T^*_p\Mcal{M})^\perp}
\newcommand{\wavetube}{wavetube }

\newcommand{\NewtonG}{G_N}

\newcommand{\Kk}{{\mathfrak{K}}}

\newcommand{\Refind}{\Mcal{N}}

\newcommand{\Ar}{\Mcal{A}}
\newcommand{\Pe}{\Mcal{P}}

\newcommand{\EM}{electromagnetic }
\newcommand{\SI}{Sagnac interferometer }

\newcommand{\TorusCarc}{\bfC}
\newcommand{\TorusC}{\bfGamma}

\newcommand{\Avtau}{\bar{\tau}}
\newcommand{\Sgn}{\text{sgn}}
\newcommand{\Cc}[1]{\overline{#1}}

\newcounter{subfig}
\newcommand{\Subfignum}{\renewcommand{\thefigure}{\arabic{figure}\alph{subfig}}\setcounter{subfig}{1}}
\newcommand{\Normfignum}{\renewcommand{\thefigure}{\arabic{figure}}}
\newcommand{\Stepfigc}{\addtocounter{figure}{-1}\addtocounter{subfig}{1}}

\begin{document}
%--------------------------

\title{\bfseries Twisted Electromagnetic Modes and Sagnac Ring-Lasers}
\author{\bfseries David A. Burton\thanks{Department of Physics,
Lancaster University, UK (email : d.burton@lancaster.ac.uk)}\\
\bfseries Adam Noble \thanks{Department of Physics,
Lancaster University, UK (email : a.noble1@lancaster.ac.uk)}\\
\bfseries Robin W. Tucker\thanks{Department of Physics, Lancaster
University, UK (email : r.tucker@lancaster.ac.uk)}\\
\bfseries David L. Wiltshire\thanks{Department of Physics and
Astronomy, University of Canterbury, Private Bag 4800, Christchurch, NZ (email :
david.wiltshire@canterbury.ac.nz)}}

\maketitle

\begin{abstract}

A new approximation scheme, designed to solve the covariant
Maxwell equations inside a rotating hollow slender conducting
cavity (modelling a ring-laser), is constructed. It is shown that
for well-defined conditions there exist TE and TM modes with
respect to the longitudinal axis of the cavity. A twisted mode
spectrum is found to depend on the integrated Frenet torsion of
the cavity and this in turn may affect the Sagnac beat frequency
induced by a non-zero rotation of the cavity. The analysis is
motivated by attempts to use ring-lasers to measure terrestrial
gravito-magnetism or the Lense-Thirring effect produced by the
rotation of the Earth.

\end{abstract}

\section{Introduction}
%%%%%%%%
In 1893 and 1897 Lodge argued that a device using \EM fields should in
principle be able to detect angular acceleration by the
interference of light. Sixteen years later Sagnac \cite{sagnac:1913}
observed such an effect and realised the potential for light interferometers as
accelerometers. The continued development of such instruments and
their careful exploitation by Michelson and others was pivotal in
the establishment of our current world view of classical physics.
Sagnac recognised that his observations could open up new avenues
in the development of accelerometers, an insight that has been
dramatically verified. Today the ring-laser is a cornerstone of
most inertial guidance systems and optoelectronics is the backbone
of the communications industry (see
\cite{post:1967, anderson_et_al:1994, stedman:1997} for historical accounts of
ring-laser development).

The basic principles leading to the Sagnac effect are well
understood in broad outline. As with all interferometers, a guided
wave phenomenon (such as that experienced by electromagnetic
fields in a wave guide) is used to distinguish different
propagation paths in spacetime. The traditional \EM ``passive'' type
of \SI involves splitting a light beam into two coherent
components, forcing them to travel different paths in spacetime to
an event where they combine to produce an interference pattern.
The nature of the interference fringes can then be correlated with
the difference in the optical paths. In such a device the
frequency of the light employed is controlled externally and the
circuits in spacetime are often determined by mirrors or
continuous dielectric fibres.

With the discovery of the laser it was found that the \EM modes of
a closed tubular cavity containing an active lasing medium (excited
initially by an external RF field) could be made to produce an
interference pattern that varied with the state of rotation of the
whole apparatus. Such an instrument has been labelled an ``active''
device to distinguish it from the traditional passive
Sagnac interferometer. Certain modes of a non-rotating cavity can be made
to form standing waves. However, the co- and counter-propagating modes in a
rotating lasing cavity acquire different resonant frequencies
leading to a beat-frequency as observed by an on-board probe.
%Surprisingly, to a first approximation this
%beat-frequency is the same as that observed in a passive
%interferometer operating with light of the same frequency as that
%resonating in the non-rotating active interferometer.
In 1963 Sperry-Rand recognised that the active ring-laser could be used as
an inertial-guidance device. Today such miniturised
interferometers are used routinely in civil and military
applications.

Considerable increase in sensitivity can be achieved using larger
ring-lasers. Insulated from thermal and seismic noise,
sensitivities of $10^{-8}$ are routinely obtained in the UG1
Sagnac interferometer housed in the Cashmere Cavern in Christchurch, New
Zealand. With an effective area of $367.6\w{m}^2$ and perimeter $77 \w{m}$
it operates with a
He-Ne laser at $474 \w{THz}$ and an intracavity power of $50 \w{mW}$ at each
mirror.
%%%% This is the UG-I laser!
%It can measure a Sagnac beat frequency of $70.64 Hz$ to
%within $\pm 5\mu Hz$ or to approximately $1$ part in $1020$ of the
%lasing frequency.
High resolution can be achieved by pattern matching to periodic
variation of the Sagnac frequency induced by rotation of the
Earth. Using matching based on geophysical modelling, detection of
lunar perturbations of the Earth's rotation and the amplitudes
of the Oppolzer modes have recently been reported in
\cite{schreiber_et_al:2004}. The instantaneous direction of the
Earth's rotation axis has been measured to a precision of $1$ part
in $10^{8}$ when averaged over a time interval of several hours.

The enhanced sensitivity of such large ring-lasers owes much to
the significant improvement of mirror design in recent years. It
is imperative for the maintenance of a Sagnac beat signal that
pollution of the resonant state by competing modes is inhibited.
This requires considerable engineering skill, the use of thermally
inert materials and mirrors that can maintain the dominant cavity
resonant mode over long periods of time.

With such advances in technology it is natural to seek the
sensitivities that are necessary to discriminate the effects of
non-Newtonian gravitation on the behaviour of light \cite{LTR}.
The most important of these is the gravito-magnetic field of the
rotating Earth, predicted by Einstein's description of
gravitation.

However, even crude estimates of the effects of terrestrial
gravito-magnetism show that a number of competing effects that may
modify the classical Sagnac effect (due to the rotation of the
apparatus) need careful consideration. Such effects are often
dependent on the structure of the interferometer and so invite one
to contemplate alternative designs. Although, in general, one
expects that the sensitivity of an active ring-laser will increase
with its spatial dimensions one may also enquire about the
dependence of sensitivity on its geometry and spatial shape. In
order to discriminate gravito-magnetic effects it then becomes
necessary to understand in some detail how the geometry and shape
of the interferometer affect the resonant mode structure of the
cavity. In this way one can hope to discriminate between a number
of features that compete to modify the classical Sagnac frequency.

The effect of gravitation on the Sagnac beat frequency is usually
discussed in terms of the high frequency ray-optic description of
light. In this way one can simply relate the interference of co-
and counter-propagating \EM modes to the proper time difference
between events that terminate their ray histories in spacetime
\cite{ashtekar_et_al:1975}. While such an analysis can give valuable
insights into the leading terms in an asymptotic expansion in the
lasing frequency it cannot address the effects due to the vector
nature of the \EM modes and the effects of the infinity of
competing modes possessed by the lasing cavity.

An exact analysis of the wave nature of light in an active Sagnac
cavity is a non-trivial mathematical problem that has to our
knowledge never been fully considered. The problem is complicated
by the fact that the harmonic \EM modes rarely split into simpler
transverse electric (TE) and transverse magnetic (TM) modes
that greatly simplify the analysis in most
simply-connected axially symmetric cavity geometries. Such a
separation relies on a number of factors such as the shape of the
entire cavity in space, the homogeneity of the lasing medium, the
nature of the acceleration of the cavity and the presence of even
a weak non-Newtonian gravitational field.

In order to accommodate these complexities a viable approximation
is mandatory. Here the problem is approached in terms of an
expansion scheme determined by scales related to the geometry of
the cavity. Thus slender cavities are considered where the ratio
of the transverse dimension to the cavity length is small. Such
cavities will be called ``wavetubes''. In order to confidently
predict near-Earth general relativistic contributions to the
active Sagnac beat frequency a careful \EM mode analysis of the
cavity must be considered. This can be done in terms of the
covariant Maxwell equations
\cite{anderson_et_al:1969, ryon_et_al:1970}
for moving media \cite{neutze:1995} on a background
spacetime determined by Einstein's equations for gravitation. The
procedure will exploit a local system of coordinates adapted to
the geometry of a wavetube. Such a wavetube can be characterised
in terms of the locus in space of the centre of each of its
transverse cross-sections. This locus is regarded as a closed
space curve with possible non-zero Frenet torsion and curvature.
The approximation scheme will be developed for a wavetube with
central locus having local curvature departing slightly from that
of a fixed plane circle.

In sections 2 and 3 it is shown how to construct TE and TM type solutions
to the covariant Maxwell equations for a particular class of
coframes. These include the fields in rotating cavities containing a
homogenous, isotropic and dispersionless medium in the
presence of terrestrial gravito-magnetism. The language of
exterior differential forms is used throughout since it offers the
most succinct formulation for the field equations and makes clear
the role of the spacetime metric and local coframe that feature in
the subsequent computational scheme. The relation to the inertial
3-d formulation can be found in \cite{benn_et_al:1987} and references therein.
In section 4 a particular coordinate system is introduced that,
together with a particular Frenet frame, encodes the geometry of
the wavetube and permits one to introduce dimensionless parameters
that control the nature of the approximation scheme. In these
coordinates differences between non-inertial motion and
gravito-magnetic effects are made explicit in sections 5.1 and
5.2. A complete mode analysis of the rotating wavetube is given in
section 6 and the effects of multi-mode excitation and the description
of Sagnac beats is explored in section 7. In particular
attention is drawn to the effects of non-planarity of the wavetube on the
beat frequency by reference to a particular geometry generated by
the structure of a torus-knot.

\section{Exact solutions to the vacuum Maxwell equations}
%%%%%%%%
\label{section:exact_EM_solutions}
To accommodate the ideas above it is first shown how a class of
spacetime metrics that admit local orthonormal coframes with
particular properties are sufficient to construct TE and TM
type \EM modes relative to such a coframe. The language of
exterior forms is used throughout\footnote{Lowercase indices at
the beginning of the Latin alphabet $\Set{a,b,c,\dots}$ range over
the integers $\Set{0,1,2,3}$, indices in the middle of the Latin
alphabet $\Set{j,k,l,\dots}$ range over the integers $\Set{2,3}$
and Greek indices $\Set{\alpha,\beta,\gamma,\dots}$ range over
$\Set{0,1}$. Expressions involving repeated indices imply use of
the Einstein summation convention.} since this greatly
facilitates the discussion and leads in a direct manner to an
explicit construction of the fields below. The hollow wavetube is
supposed to be constructed of a perfectly conducting material
which in this section contains no lasing medium.

Let $(\Mcal{M},g)$ be a spacetime and let
$\Set{e^0,e^1,e^2,e^3}$ be a local orthonormal coframe with
respect to the metric tensor
\begin{equation*}
g = -e^0\Tensor e^0 + e^1\Tensor e^1 + e^2\Tensor e^2 + e^3\Tensor e^3
\end{equation*}
with dual frame $\Set{X_0,X_1,X_2,X_3}$,
\begin{equation}
e^a(X_b) = \delta^a_b,
\end{equation}
and orientation (volume form) defined by the Hodge map $\Hodge$ with
\begin{equation*}
\Hodge 1 = e^{0}\wedge e^{1}\wedge e^{2}\wedge e^{3}.
\end{equation*}
The inverse metric tensor $\Invg$ is
\begin{equation*}
\Invg = -X_0\Tensor X_0 + X_1\Tensor X_1 + X_2\Tensor X_2 + X_3\Tensor
 X_3.
\end{equation*}
Define the split of the cotangent space
$T_p^*\Mcal{M}=\Tstarpar\oplus\Tstarperp$ with $\Tstarpar$
spanned by $\Set{e^0,e^1}$ and $\Tstarperp$ spanned
by $\Set{e^2,e^3}$ at $p\in\Mcal{M}$.
These subspaces inherit Hodge maps with $\ParHodge
1=e^0\wedge e^1$ and $\PerpHodge 1=e^2\wedge e^3$ so
\begin{equation*}
\Hodge 1 = \ParHodge 1 \wedge \PerpHodge 1.
\end{equation*}
Now restrict to spacetime metrics $g$ that admit local orthonormal coframes
satisfying the conditions
\begin{gather}
\notag
de^\alpha = 0,\\
\label{coframe_assumptions}
d\PerpHodge 1 = 0.
\end{gather}
These conditions permit a viable approximation scheme leading to a
tractable decomposition of electromagnetic modes in a wide context to
be discussed below.

A differential form $\mu$ (vector field $X$) on $\Mcal{M}$ that satisfies
$\Intder{X_j}\mu=0$ ($e^j(X)=0$) for all $j$
will be called \emph{longitudinal}\footnote{$\Intder{X}$ is the
interior derivative on forms with respect to the vector field $X$.}.
A \emph{transverse}
differential form $\nu$ (vector field $Y$) will satisfy
$\Intder{X_\alpha}\nu=0$ ($e^\alpha(Y)=0$) for all
$\alpha$.

Let the $2$-form $F$ be a solution to the
vacuum Maxwell equations
\begin{gather}
\notag
dF=0,\\
\label{Maxwell_dF_dstarF}
d\Hodge F=0
\end{gather}
in a spacetime region bounded
by a perfectly conducting \wavetube hypersurface
$\Mcal{B}\equiv\Set{p\in\Mcal{M}:f(p)=0, df\,\,\text{timelike}}$
and subject to the boundary condition
\begin{equation}
\label{Fresnel_BC}
(df \wedge F)(p) = 0 \Bcon.
\end{equation}
The \wavetube interior on spacetime is topologically $\BR \cross
S^1\cross \BD$ where $\BD$ is a $2$-disc.
Solutions to Maxwell's equations are sought by adapting the above coframe
to the \wavetube geometry. Furthermore, we shall suppose that
(\ref{Fresnel_BC}) must be satisfied with $df(p)$,
for all $p\in\Mcal{B}$, a transverse $1$-form. Since $de^\alpha=0$ the pair
$\Set{e^0,e^1}$ are, by Frobenius' theorem, normal $1$-forms to a
local foliation $\Mcal{F}$ of $\Mcal{M}$. The leaves of $\Mcal{F}$ contain the \wavetube
cross-sections, each of whose intersection with $\Mcal{B}$ is the image of a closed curve.

Locally, we seek separable
propagating solutions of the form $F=\exp(iW)\gamma$ for some
$2$-form $\gamma$ where
\begin{equation*}
dW = \xi_\alpha e^{\alpha} \equiv \xi \neq 0
\end{equation*}
for constant components $\Set{\xi_0,\xi_1}$. We decompose $\gamma$
into longitudinal and transverse parts
\begin{gather*}
\gamma = \Phi\ParHodge 1 + \Psi\PerpHodge 1 +
e^\alpha\wedge\Kkappa_\alpha,\\
\Kkappa_\alpha(X_\beta)=0,\\
d\Phi(X_\alpha)=0,\\
d\Psi(X_\alpha)=0,\\
\Intder{X_\alpha}d\Kkappa_\beta=0.
\end{gather*}
Thus $\Set{d\Phi,d\Psi,\Kkappa_0,\Kkappa_1}$ are transverse
$1$-forms and $\Set{d\Kkappa_0,d\Kkappa_1}$ are transverse $2$-forms.
Using the identity
\begin{equation*}
\Hodge(\alpha\wedge \Mdual{X}) = \Intder{X}\Hodge\alpha,
\end{equation*}
where $\Mdual{X}\equiv g(X,-)$ for any vector field $X$ on $\Mcal{M}$
and $\alpha$ is any $p$-form on $\Mcal{M}$ it can be shown that
\begin{gather}
\label{Hodge_as_ParHodge_PerpHodge}
\Hodge(\mu\wedge\nu) = (-1)^{(2-p)q}
\ParHodge\mu\wedge\PerpHodge\nu
\end{gather}
where $\mu$ is a longitudinal $p$-form, $\nu$ is a transverse
$q$-form and
\begin{gather*}
\ParHodge(\alpha\wedge\Mdual{X}) = \Intder{X}\ParHodge\alpha,\\
\PerpHodge(\alpha\wedge\Mdual{X}) = \Intder{X}\PerpHodge\alpha.
\end{gather*}
Inserting the expression for $F$ into Maxwell's equations
(\ref{Maxwell_dF_dstarF}) gives
\begin{gather*}
d\Phi\wedge\ParHodge 1 - e^\alpha\wedge d\Kkappa_\alpha + i \xi\wedge
(\Psi\PerpHodge 1 + e^\alpha\wedge\Kkappa_\alpha)=0,\\
d\Psi\wedge\ParHodge 1 + \ParHodge e^\alpha\wedge
d\PerpHodge\Kkappa_\alpha - i \xi\wedge(\Phi\PerpHodge 1 + \ParHodge
e^\alpha\wedge\PerpHodge \Kkappa_\alpha)=0
\end{gather*}
where (\ref{Hodge_as_ParHodge_PerpHodge}) has been used.
The form structure indicates that the above pair splits into
\begin{gather}
\label{split_maxwell_dphi}
d\Phi\wedge\ParHodge 1 + i \xi\wedge e^\alpha\wedge\Kkappa_\alpha=0,\\
\label{split_maxwell_ealpha}
e^\alpha\wedge d\Kkappa_\alpha - i \xi\wedge\Psi\PerpHodge 1=0,\\
\label{split_maxwell_dpsi}
d\Psi\wedge\ParHodge 1 - i \xi\wedge\ParHodge e^\alpha\wedge\PerpHodge
\Kkappa_\alpha=0,\\
\label{split_maxwell_parhodgeealpha}
\ParHodge e^\alpha\wedge d\PerpHodge\Kkappa_\alpha - i\xi\wedge\Phi\PerpHodge 1=0.
\end{gather}
Acting with $\Hodge$ on (\ref{split_maxwell_dphi}) and
(\ref{split_maxwell_dpsi}) and using
(\ref{Hodge_as_ParHodge_PerpHodge}) yields
\begin{gather}
\label{phi_Kkappa_related}
\Kkappa_{(\ParHodge \xi)} = -i d\Phi,\\
\label{psi_Kkappa_related}
\Kkappa_{(\xi)} = i\PerpHodge d\Psi
\end{gather}
where $\Kkappa_{(\xi)} =
\xi^\alpha \Kkappa_\alpha$ and $\Kkappa_{(\ParHodge\xi)} =
(\ParHodge\xi)^\alpha \Kkappa_\alpha$. Acting with
$\Intder{\Mdual{\xi}}$, where $\Mdual{\xi}=\Invg(\xi,-)$,
on equation (\ref{split_maxwell_ealpha}) yields
\begin{equation}
\label{dKkappa_psi_related}
d\Kkappa_{(\xi)} = i \xi^2\Psi\PerpHodge 1
\end{equation}
where $\xi^2 \equiv g^{-1}(\xi,\xi)$. Similarly,
(\ref{split_maxwell_parhodgeealpha}) leads to
\begin{equation}
\label{dKkappa_phi_related}
d\PerpHodge\Kkappa_{(\ParHodge \xi)} = -i \xi^2 \Phi\PerpHodge 1.
\end{equation}
Thus, inserting (\ref{phi_Kkappa_related}) in (\ref{dKkappa_phi_related})
and (\ref{psi_Kkappa_related}) in (\ref{dKkappa_psi_related}) yields
\begin{gather}
\label{equation_for_phi}
d\PerpHodge d\Phi = \xi^2\Phi\PerpHodge 1,\\
\label{equation_for_psi}
d\PerpHodge d\Psi = \xi^2\Psi\PerpHodge 1.
\end{gather}
The Helmholtz equations (\ref{equation_for_phi}) and (\ref{equation_for_psi})
determine $\Phi$ and $\Psi$ subject to (\ref{Fresnel_BC}).
The boundary condition (\ref{Fresnel_BC}) may be expanded
\begin{equation*}
df(p) \wedge (\Phi\ParHodge 1 + e^\alpha\wedge\Kkappa_\alpha)(p) = 0\Bcon
\end{equation*}
implying
\begin{gather}
\label{phi_boundary_condition}
\Phi(p) = 0\Bcon,\\
\label{Kkappa_boundary_condition}
(df \wedge e^\alpha\wedge\Kkappa_\alpha)(p) = 0\Bcon.
\end{gather}

In general there are three possibilities for the metric norm
$\xi^2=g^{-1}(\xi,\xi)$ of $\xi$ :
\newcounter{c1}
\begin{list}{$\bullet$}
{\usecounter{c1}}
\item $\xi^2=0$.

\textbf{Triviality of $\Phi$}\\
Let $S$ be a $2$-chain for which the image of
$\Bdry S$ is in $\Mcal{B}$ and $(\Bdry S)^* df = 0$ and note
\begin{equation}
\label{integration_Phi}
\begin{split}
\int\limits_S d\Cc{\Phi}\wedge\PerpHodge d\Phi &=
\int\limits_{\Bdry S}\Cc{\Phi}\PerpHodge d\Phi -
\int\limits_{S}\Cc{\Phi}d\PerpHodge d\Phi\\
&= -\xi^2 \int\limits_S |\Phi|^2\PerpHodge 1
\end{split}
\end{equation}
using (\ref{equation_for_phi}) and (\ref{phi_boundary_condition})
where $\Cc{\Phi}$ is the complex-conjugate of $\Phi$. Since
\begin{equation}
\label{positivity_ParHodge}
\PerpHodge^{-1}(\beta\wedge\PerpHodge\beta) \ge 0
\end{equation}
where $\beta$ is any transverse $p$-form and, by hypothesis, $\xi^2=0$
it follows from (\ref{integration_Phi}) that $d\Phi=0$ and so, using
(\ref{phi_boundary_condition}), $\Phi=0$.

\textbf{Triviality of $\Psi$ and $F$}\\
Since $\xi$ is null $\ParHodge \xi$ is proportional to $\xi$
and so (\ref{phi_Kkappa_related}) and
(\ref{psi_Kkappa_related}) with $\Phi=0$ imply that $\Kkappa_{(\xi)}$
vanishes and $\Psi$ is
constant. Let $\zeta=\zeta_\alpha e^\alpha$ be the null longitudinal $1$-form with
constant components $\zeta_\alpha$ that satisfies $g^{-1}(\xi,\zeta) =
1$. Acting with $\Intder{\Mdual{\zeta}}$ on
(\ref{split_maxwell_parhodgeealpha}) and
(\ref{split_maxwell_ealpha}) and noting $\ParHodge\zeta = \varepsilon\zeta$ where
$|\varepsilon|=1$ it is found that
\begin{gather}
\label{div_Kkappa_l}
d\PerpHodge\Kkappa_{(\zeta)} = 0,\\
\label{dKkappa_l_Psi_related}
d\Kkappa_{(\zeta)} = i\Psi\PerpHodge 1.
\end{gather}
Since $\Kkappa_{(\xi)}$ vanishes $e^\alpha\wedge\Kkappa_\alpha =
(\xi^\alpha \zeta + \zeta^\alpha \xi) \wedge\Kkappa_\alpha =
\xi\wedge\Kkappa_{(\zeta)}$ where $\xi^\alpha=g^{-1}(e^\alpha,\xi)$
and $\zeta^\alpha=g^{-1}(e^\alpha,\zeta)$. The
differential form $df\wedge\Kkappa_{(\zeta)}$ is transverse and so
\begin{equation}
\label{Kkappa_l_boundary_condition}
(df \wedge\Kkappa_{(\zeta)})(p) = 0 \Bcon
\end{equation}
follows from (\ref{Kkappa_boundary_condition}).
By integrating (\ref{dKkappa_l_Psi_related}) over the $2$-chain $S$
introduced earlier, noting that (\ref{Kkappa_l_boundary_condition})
implies $(\Bdry S)^*\Kkappa_{(\zeta)}=0$ and recalling that
$\Psi$ is constant, it follows that $\Psi=0$ and
\begin{equation*}
d\Kkappa_{(\zeta)} = 0.
\end{equation*}
Cross-sections of the \wavetube interior are
simply-connected and so by the Poincar\'e lemma \cite{abraham_et_al:1988}
\begin{equation}
\label{exactness_of_Kkappa_zeta}
\Kkappa_{(\zeta)} = d\varphi.
\end{equation}
It follows from (\ref{div_Kkappa_l}) and (\ref{Kkappa_l_boundary_condition})
that $\varphi$ satisfies
\begin{equation}
\label{Laplace_in_null_case}
d\PerpHodge d\varphi = 0
\end{equation}
subject to the Dirichlet boundary condition
\begin{equation}
\label{Laplace_in_null_case_boundary_cond}
\varphi(p) = \varphi_0 \Bcon
\end{equation}
where $\varphi_0$ is a complex constant. The solution to
(\ref{Laplace_in_null_case}) and
(\ref{Laplace_in_null_case_boundary_cond}) is
$\varphi=\varphi_0$ and so, using (\ref{exactness_of_Kkappa_zeta}),
$\Kkappa_{(\zeta)}=0$ and
\begin{equation*}
F=0.
\end{equation*}

\item $\xi^2>0$.

\textbf{Triviality of $\Phi$}\\
Using (\ref{integration_Phi}) and (\ref{positivity_ParHodge}) it can
be seen that $\Phi$ vanishes since, by hypothesis, $\xi^2>0$.

\textbf{Triviality of $\Psi$ and $F$}\\
Introduce the spacelike
normalized $1$-form $m=\xi/\sqrt{\xi^2}$ and write
\begin{equation*}
\begin{split}
\Kkappa_\alpha &= m_\alpha\Kkappa_{(m)} - (\ParHodge
m)_\alpha\Kkappa_{(\ParHodge m)}\\
&= m_\alpha\Kkappa_{(m)}
\end{split}
\end{equation*}
where $m_\alpha=m(X_\alpha)$ and $(\ParHodge m)_\alpha=(\ParHodge
m)(X_\alpha)$ and $\Kkappa_{(\ParHodge m)}$ vanishes by
(\ref{phi_Kkappa_related}) because $\Phi=0$.
Thus (\ref{Kkappa_boundary_condition}) yields
\begin{equation*}
(df \wedge \xi \wedge\Kkappa_{(\xi)})(p) = 0 \Bcon.
\end{equation*}
Since $\xi$ is longitudinal and $df\wedge\PerpHodge d\Psi$
is transverse (\ref{psi_Kkappa_related}) gives the Neumann boundary condition
\begin{equation}
\label{psi_boundary_condition}
(df \wedge\PerpHodge d\Psi)(p) = 0\Bcon.
\end{equation}
The simple-connectedness of the \wavetube cross-sections
implies that the eigenvalues $\xi^2$ in (\ref{equation_for_psi}) associated with
the Neumann boundary condition (\ref{psi_boundary_condition}) are
negative. To prove this an argument similar to that used in the
Dirichlet case is employed. Let $S$ be the $2$-chain introduced
earlier and note that (\ref{psi_boundary_condition}) implies
$\PerpHodge d\Psi = hdf$, where $h$ is a $0$-form, and
\begin{equation}
\label{psi_boundary_condition_S}
(\Bdry S)^*(\PerpHodge d\Psi)=0
\end{equation}
since $(\Bdry S)^*df=0$. Thus
\begin{equation}
\label{integration_Psi}
\begin{split}
\int\limits_S d\Cc{\Psi}\wedge\PerpHodge d\Psi &=
\int\limits_{\Bdry S}\Cc{\Psi}\PerpHodge d\Psi -
\int\limits_{S}\Cc{\Psi}d\PerpHodge d\Psi\\
&= -\xi^2 \int\limits_S |\Psi|^2\PerpHodge 1
\end{split}
\end{equation}
using (\ref{equation_for_psi}) and (\ref{psi_boundary_condition_S})
where $\Cc{\Psi}$ is the complex-conjugate of $\Psi$. By hypothesis
$\xi^2>0$ and using (\ref{positivity_ParHodge}) it follows that
$\Psi=0$. Therefore
\begin{equation*}
F=0.
\end{equation*}

\item $\xi^2<0$.

This is the only situation that leads to a non-trivial
expression for $F$. Introduce the timelike
normalized $1$-form $m= \xi/|\xi|$ and write
\begin{equation*}
\begin{split}
\Kkappa_\alpha &= -m_\alpha\Kkappa_{(m)} + (\ParHodge
m)_\alpha \Kkappa_{(\ParHodge m)}\\
&= -i\PerpHodge d\Psi \frac{\xi_\alpha}{|\xi|^2} - i d\Phi \frac{(\ParHodge
\xi)_\alpha}{|\xi|^2}
\end{split}
\end{equation*}
where $|\xi|\equiv \sqrt{-\xi^2}$ and
(\ref{phi_Kkappa_related}) and (\ref{psi_Kkappa_related}) have been used.
Inserting the above equation for
$\Kkappa_\alpha$ into the boundary condition
(\ref{Kkappa_boundary_condition}) yields
\begin{equation*}
\biggl[df\wedge \biggl(\frac{\xi}{|\xi|^2} \wedge\PerpHodge d\Psi + \frac{\ParHodge
\xi}{|\xi|^2} \wedge d\Phi \biggr)\biggr](p) = 0 \Bcon.
\end{equation*}
However, (\ref{phi_boundary_condition}) means that $d\Phi$ on
$\Mcal{B}$ is proportional to $df$
and so again (\ref{psi_boundary_condition}) is obtained.
\end{list}

Expressed entirely in terms of $\Set{\Phi,\Psi,W}$ the Maxwell
2-form $F$ can be written $F=F_{TM}+F_{TE}$ where
\begin{gather}
\label{TM_F_in_terms_of_Phi}
F_{TM} = \biggl[ \Phi\ParHodge 1 - i\frac{\ParHodge
dW_{TM}}{|dW_{TM}|^2} \wedge d\Phi\biggr]\exp(iW_{TM}),\\
\label{TE_F_in_terms_of_Psi}
F_{TE} = \biggl[ \Psi\PerpHodge 1 - i\frac{dW_{TE}}{|dW_{TE}|^2}
\wedge \PerpHodge d\Psi\biggr]\exp(iW_{TE}).
\end{gather}
The Helmholtz equations for $\Phi$ and $\Psi$ are
\begin{gather*}
d\PerpHodge d\Phi = -|dW_{TM}|^2\Phi\PerpHodge 1,\\
d\PerpHodge d\Psi = -|dW_{TE}|^2\Psi\PerpHodge 1
\end{gather*}
solved subject to the boundary conditions
\begin{gather*}
\Phi(p) = 0\Bcon,\\
(df \wedge\PerpHodge d\Psi)(p) = 0\Bcon
\end{gather*}
where $dW_{TM}=\xi^{TM}_\alpha e^\alpha$ and $dW_{TE}=\xi^{TE}_\alpha
e^\alpha$ are timelike. The constants
$\Set{\xi^{TM}_\alpha, \xi^{TE}_\alpha}$ are
determined by the appropriate boundary conditions and, in general,
$dW_{TE}\neq dW_{TM}$.
\section{Exact solutions to the Maxwell equations in a medium}
%%%%%%%%
To excite electromagnetic modes in a wavetube one may fill it
with a gas such as a helium-neon mixture and induce it to lase
with an external RF field. The presence of a lasing medium
requires that one take into account its electrical and magnetic
susceptibility. The electromagnetic field in a material medium is
described by a pair of 2-forms $\{F,G\}$ on spacetime that satisfy the
equations:
\begin{gather*}
dF=0,\\
d\Hodge G=j
\end{gather*}
where $j$ is a source 3-form.

For a given $j$, the 2-form $G$ must be related to $F$ in order to
have a closed system. This is usually done by relating their
components relative to decompositions of the form:
\begin{gather*}
F=E\wedge\Mdual{V} + B,\\
G=D\wedge\Mdual{V} + H
\end{gather*}
where the timelike vector $V$ is normalised with $g(V,V) =-1$ and the
forms $\{E,B,D,H\}$ are all annihilated by $\Intder{V}$.
With a spacetime metric tensor $g$, having physical
dimensions of length squared, a coherent dimensional scheme arises
with the aid of the permittivity constant $\epsilon_0$ and
permeability $\mu_0$ of free space satisfying
$\epsilon_0\,\mu_0=\frac{1}{c^2}$ where $c$ is the speed of light in
vacuum. In terms of conventional MKS
components of electric and magnetic fields one has in any
$g$-orthonormal coframe $\{f^0=-\Mdual{V},f^1,f^2,f^3\}$ adapted
to $V$:
\begin{gather*}
E:=\epsilon_0 E^{MKS}=\epsilon_0(E_1^{MKS} f^1 + E_2^{MKS} f^2 +E_3^{MKS} f^3 ),\\
B:=\frac{1}{\mu_0} B^{MKS}=\frac{1}{\mu_0}(B_1^{MKS} f^2 \wedge f^3 +
B_2^{MKS} f^3 \wedge f^1 + B_3^{MKS} f^1 \wedge f^2).
\end{gather*}
For $G$ one has $D=D^{MKS}$ and $H=H^{MKS}$. For a linear
isotropic homogeneous dispersionless non-conducting medium with
relative permittivity $\epsilon_r$ and relative permeability $\mu_r$ one has the
simple constitutive relations $D=\epsilon_r E$ and $H= B/\mu_r$ or
\begin{equation*}
G=(\epsilon_r - \frac{1}{\mu_r}) \Intder{V} F \wedge \Mdual{V} +
\frac{1}{\mu_r} F.
\end{equation*}
Such a constitutive relation assumes that the dimensionless
quantities $\epsilon_r$ and $\mu_r$ are frame-independent constant scalars
on spacetime. The refractive index $\Refind$ of the medium is
\begin{equation*}
\Refind = \sqrt{\epsilon_r\mu_r}.
\end{equation*}

Propagating solutions to the source-free Maxwell equations
\begin{gather}
\notag
dF=0,\\
\label{Source_free_Maxwell_medium}
d\Hodge G=0
\end{gather}
that satisfy the boundary condition
\begin{equation*}
(df \wedge F)(p) = 0 \Bcon
\end{equation*}
at the wavetube surface $\Mcal{B}$,
\begin{equation*}
\Mcal{B}=\Set{p\in\Mcal{M}:f(p)=0}, df\,\,\text{timelike},
\end{equation*}
have a similar structure to (\ref{TM_F_in_terms_of_Phi}) and
(\ref{TE_F_in_terms_of_Psi}) when $V$ has the form
\begin{equation*}
V = V^\alpha X_\alpha,
\end{equation*}
where $\{V^\alpha\}$ are constant and $\{X_\alpha\}$ belong to the
frame $\{X_a\}$ dual to the coframe $\{e^a\}$ satisfying the
conditions (\ref{coframe_assumptions}).

It can be shown that $F_{TM}$ and
$F_{TE}$ are solutions to (\ref{Source_free_Maxwell_medium}) where
\begin{gather}
\label{TM_F_medium_in_terms_of_Phi}
F_{TM} = \biggl[ \Phi\ParHodge 1 - i\frac{\ParHodge
\beta_{TM}}{|\alpha_{TM}|^2} \wedge d\Phi\biggr]\exp(iW_{TM}),\\
F_{TE} = \biggl[ \Psi\PerpHodge 1 - i\frac{dW_{TE}}{|\alpha_{TE}|^2}
\wedge \PerpHodge d\Psi\biggr]\exp(iW_{TE})
\label{TE_F_medium_in_terms_of_Psi}
\end{gather}
with
\begin{gather}
\notag
\alpha_{TM} = (1-\Refind)dW_{TM}(V)\Mdual{V} + dW_{TM},\\
\notag
\alpha_{TE} = (1-\Refind)dW_{TE}(V)\Mdual{V} + dW_{TE},\\
\label{alpha_T_beta_T_explicit}
\beta_{TM} = (1-\Refind^2)dW_{TM}(V)\Mdual{V} + dW_{TM}.
\end{gather}
The Helmholtz equations for $\Phi$ and $\Psi$ are
\begin{gather*}
d\PerpHodge d\Phi = -|\alpha_{TM}|^2\Phi\PerpHodge 1,\\
d\PerpHodge d\Psi = -|\alpha_{TE}|^2\Psi\PerpHodge 1
\end{gather*}
and are solved subject to the boundary conditions
\begin{gather*}
\Phi(p) = 0\Bcon,\\
(df \wedge\PerpHodge d\Psi)(p) = 0\Bcon.
\end{gather*}
As before $dW_{TM}=\xi^{TM}_\alpha e^\alpha$ and $dW_{TE}=\xi^{TE}_\alpha
e^\alpha$ where $\Set{\xi^{TM}_\alpha}$ and $\Set{\xi^{TE}_\alpha}$
are constant.
\section{Twisted ring cavities rotating in space}
%%%%%%%%
\subsection{Moving space curves}
%%%%%%%%%%%
A moving space curve in Euclidean $\BR^3$ may be represented by a
$\BR^3$-valued mapping $(s,t)\mapsto \bfC(s,t)$. Suppose that $s$
parameterises arc length and $t$ is a time variable. The
instantaneous geometry of the space curve may be described by
three mutually orthogonal unit vectors $\Set{\bft,\bfn,\bfb}$ on the
image of $\bfC$ at each time $t_0$. For a Frenet frame with $\bft$
tangent to $\bfC(s,t_0)$ one has:
\begin{gather}
\notag
\bfC^\prime = \bft,\\
\notag
\bft^\prime=\kappa\bfn,\\
\notag
\bfn^\prime=-\kappa\bft+\tau\bfb,\\
\label{Frenet_Serret_equations}
\bfb^\prime=-\tau\bfn
\end{gather}
where $f^\prime$ denotes the derivative of $f$ with respect to $s$
and $\kappa(s,t_0)$ and $\tau(s,t_0)$ denote the curvature and
torsion of $\bfC(s,t_0)$ respectively. We shall restrict our
discussion to space curves that admit continuous Frenet frames
along their entire length. Since the triad $(\bft,\bfn,\bfb)$
forms an orthonormal basis on the curve the time derivative
$(\dot\bft,\dot\bfn,\dot\bfb)$ is related to $(\bft,\bfn,\bfb)$ by
an anti-symmetric matrix for each $(s,t)$. We are interested in
space curves that rotate rigidly with uniform angular velocity
${\bfOm }$ without deformation. Thus $\bfOm^\prime=0,
\dot\bfOm=0$, $\dot\kappa=\dot\tau=0$ and $\dot\bfC=\bfOm\times
\bfC$.

These imply
\begin{gather}
\notag
\dot\bft=\bfOm\cross\bft,\\
\notag
\dot\bfn=\bfOm\cross\bfn,\\
\label{action2}
\dot\bfb=\bfOm\cross \bfb.
\end{gather}
In a fixed global cartesian triad $(\bfi,\bfj,\bfk)$ one may write
in cylindrical coordinates:
\begin{equation*}
\bfC=C_r\cos(C_\theta)\bfi + C_r\sin(C_\theta)\bfj + C_k\bfk.
\end{equation*}
If this triad is oriented so that $\bfOm=\Omega\bfk$ then $C_k$
and $C_r$ are independent of $t$ and $C_\theta(s,t)=C_\theta(s)+\Omega t$.

Let the vector $\bfr_{t_0}$ with Cartesian components $(y_1,y_2,y_3)$
in the frame $(\bfi,\bfj,\bfk)$,
\begin{equation*}
\bfr_{t_0}=y_1\bfi+y_2\bfj+y_3\bfk,
\end{equation*}
locate a point at the instant $t_0$.
If this point lies in a tubular neighbourhood about the space curve
$\bfC(s,t_0)$ it may also be labelled by the Frenet coordinates
$(s,x_1,x_2)$ such that:
\begin{equation*}
\bfr_{t_0}=\bfC(s,t_0)+x_1\bfn(s,t_0) + x_2\bfb(s,t_0)
\end{equation*}
Thus for points in this region one has a transformation between
the Cartesian coordinates $(y_1,y_2,y_3)$ and the Frenet coordinates
$(s, x_1,x_2)$.

When the curve moves
\begin{equation*}
d\bfr=d\bfC + dx_1\bfn+dx_2\bfb + x_1 d\bfn + x_2 d\bfb
\end{equation*}
and so for rigid motion the above gives:
\begin{equation*}
\begin{split}
d\bfr =& [\bfOm\cross \bfC + x_1 (\bfOm\cross \bfn) +
x_2(\bfOm\cross \bfb)]dt\\
&+ [1 -\kappa(s) x_1]ds\,\bft +[d x_1 - x_2 \tau(s) ds]\bfn +[d
x_2 + x_1 \tau(s) ds]\bfb\\
=& dy_1\bfi + dy_2\bfj + dy_3\bfk
\end{split}
\end{equation*}
where $dy_1=d\bfr\cdot \bfi$,
$dy_2=d\bfr\cdot \bfj$ and $dy_3=d\bfr\cdot \bfk$.
These equations imply a $t$ dependent coordinate transformation
between $(y_1,y_2,y_3)$ and $(s,x_1,x_2)$.

The following is concerned with cavities that have small circular
cross-section compared with their length and are represented by a space curve
that approximates a plane circle rotating about its axis.
To this end it is convenient to
introduce a number of dimensionless parameters and new
coordinates. First introduce polar variables $\rho$ and $\phi$ by
\begin{gather*}
x_1=\rho\cos\phi,\\
x_2=\rho\sin\phi
\end{gather*}
where $0\le\rho\le\rho_0$, $0<\phi\le 2\pi$ and $\rho_0$ is the
radius of the circular cross-section.

It is supposed that the absolute values of the curvature and torsion of the closed
curve $\bfC$ are bounded from above by $\kappa_0$ and $\tau_0$ respectively
and that the scale of $\bfC$ is given by that of a circle of
radius $a_0$. Introduce the dimensionless parameters
\begin{gather*}
\epsilon\mu_1=\frac{\Om\rho_0}{c},\\
\mu_2=\frac{\Om a_0}{c},\\
\epsilon\mu_3=\rho_0\kappa_0,\\
\mu_4=\rho_0\tau_0.
\end{gather*}
The dimensionless parameter $\epsilon$ indicates
the order of small quantities that will be used below to
approximate the metric appropriate for a cavity with small cross
section. Alternative assumptions about the assignment of $\epsilon$
to the quantities $\Set{\mu_1,\mu_2,\mu_3,\mu_4}$ will give different
approximation schemes. The above choice will be shown to give the
classical Sagnac beat frequency to first order in a \wavetube based on
a planar space curve.

The calculation will proceed for
\begin{equation*}
d\bfr\cdot d\bfr \equiv dy_1\Tensor dy_1 + dy_2\Tensor dy_2 + dy_3\Tensor dy_3
\end{equation*}
to lowest order in $\epsilon$ for a curve that approximates a
circle\footnote{Possibly covered more than once (see Figure 1).}
of radius $a_0$ rotating with uniform angular speed $\Om$ about its
axis. Introduce the scaled polar coordinates $(\Hs,\Hrho,\phi)$
and $\Hlambda$ where
\begin{gather*}
\Hlambda = \frac{ct}{\rho_0},\\
\Hs = \frac{s}{\rho_0},\\
\Hrho = \frac{\rho}{\rho_0}
\end{gather*}
and the scaled quantities
\begin{gather*}
\Hkappa(\Hs)=\frac{\kappa(s)}{\kappa_0},\\
\Htau(\Hs)=\frac{\tau(s)}{\tau_0}
\end{gather*}
and consider the class of $\bfC$ for which
\begin{gather}
\notag
[\bft\cdot(\bfk\cross\bfC)](s,t) = a_0\bigl[1 +
\epsilon\Gamma_1(\Hs,\Hlambda)\bigr],\\
\notag
[\bfn\cdot(\bfk\cross\bfC)](s,t) = a_0\epsilon\Gamma_2(\Hs,\Hlambda),\\
\label{approximate_curve_conditions}
[\bfb\cdot(\bfk\cross\bfC)](s,t) = a_0\epsilon\Gamma_3(\Hs,\Hlambda).
\end{gather}
Thus, to lowest (zeroth) order in $\epsilon$
\begin{equation*}
\frac{d\bfr}{\rho_0} = \bft d\Hs + \bfn [d\Hx_1 -\mu_4\hat\tau(\Hs) \Hx_2
d\Hs] + \bfb[d\Hx_2 +\mu_4\hat\tau(\Hs) \Hx_1 d\Hs] + \bft\mu_2 d\Hlambda.
\end{equation*}
In terms of $(\Hs,\Hrho,\phi)$ and $\Hlambda$
\begin{equation*}
\begin{split}
\frac{d\bfr\cdot d\bfr}{\rho_0^2} =& \,(d\Hs + \mu_2 d\Hlambda)\Tensor
(d\Hs + \mu_2 d\Hlambda)\\
&+\Hrho^2[d\phi+\mu_4\Htau(\Hs)
d\Hs]\Tensor[d\phi+\mu_4\Htau(\Hs) d\Hs] + d\Hrho\Tensor d\Hrho
\end{split}
\end{equation*}
and the zeroth order space curve is a circle of radius $a_0$
orthogonal to the rotation axis $\bfk$.
\section{Ring cavities on spacetime}
%%%%%%%%
We shall consider only stationary axisymmetric spacetimes. Thus
 $(\Mcal{M},g)$ admits a commuting pair of Killing vector
fields $K$ and $L$:
\begin{gather*}
\Lie{K}g = 0,\\
\Lie{L}g = 0,\\
[K,L] = 0
\end{gather*}
where $\Lie{X}$ is the Lie derivative with respect to $X$. The
vector field $K$ is timelike and future-pointing and the vector
field $L$ is spacelike and has closed orbits. Here by definition,
material points that are ``rotating''\footnote{There is no unique
definition of ``rotation'' or ``rigidity'' in general spacetimes.
Indeed there exist a
plethora of possible intrinsically defined notions of ``rotation'' that compete
for attention. Such notions are generally associated with the
alternative notions of ``local frames'' in a generic spacetime. In
the neighborhood of most events the use of timelike geodesic
coordinates yield vector fields that offer a good approximation to
the ``inertial frames'' of flat spacetime, They have natural
generalisations to Fermi-Walker charts in the neighbourhood of
arbitrary timelike worldlines. These are associated with
Fermi-Walker parallel ortho-normal frames attached to such
worldlines which may be naturally identified as constituting a
``non-rotating'' frame attached to such an observer. Fermi-Walker
frames have the virtue that they admit
 an operational definition. Such notions are local and apply to
 spacetimes with
no particular symmetries. If the spacetime has preferred timelike
global (local) vector fields, such as timelike Killing vectors,
they offer a field of preferred global (local) observers. Such
vector fields may have vorticity that contributes to a notion of
local rotation in an extended region. If there are additional
spacelike Killing fields then additional local timelike
combinations may be considered as observer fields \cite{Page}.
Preferences must be based on expediency within some operational
framework. Sagnac interferometry offers such a framework. Many
discussions of this phenomena are based on the use of ray optics
and the behaviour of photons constrained to follow null
non-geodesic paths by idealised mirrors. Such mirrors act as
``local observers'' and the particular observer field to which they
belong offers a natural frame for the analysis of the experiment.
In this article we concentrate on the electromagnetic field aspect
 and make no use of the eikonal or ray approximation. We do however
require the existence of a stationary axially symmetric spacetime
and hence the existence of the Killing vectors $K$ and $L$ above.
} follow timelike worldlines whose tangents at any event are a
linear combination $V$ of $K$ and $L$ at that event. If the
spacetime is asymptotically flat and the normalised $V$ is
timelike asymtotically it also offers a field of stationary
observers.

\subsection{Flat spacetime}
%%%%%%%%%%%
Before constructing a local coframe adapted to a twisted wavetube
rotating in a spacetime with gravitation we consider the simpler case
of Minkowski spacetime.
The metric tensor $g$ on Minkowski spacetime is
\begin{equation*}
\begin{split}
g &= -c^2 dT \otimes dT + dy_1 \otimes dy_1 + dy_2 \otimes dy_2
+dy_3 \otimes dy_3\\
&= -c^2 dT\Tensor dT + d\bfr\cdot d\bfr
\end{split}
\end{equation*}
in terms of the inertial coordinates $(T,y_1,y_2,y_3)$. It possesses
the timelike future-pointing Killing vector field $\PD/\PD T$ and the
spacelike Killing vectors $\Set{\PD/\PD y_1,\PD/\PD y_2,\PD/\PD y_3}$
and $\Set{L_1,L_2,L_3}$ where
\begin{gather*}
L_1 = y_2\frac{\PD}{\PD y_3} - y_3\frac{\PD}{\PD y_2},\\
L_2 = y_3\frac{\PD}{\PD y_1} - y_1\frac{\PD}{\PD y_3},\\
L_3 = y_1\frac{\PD}{\PD y_2} - y_2\frac{\PD}{\PD y_1}.
\end{gather*}
On the \wavetube spacetime domain $(T,y_1,y_2,y_3)$ is
related to the spacetime Frenet coordinates $(t,s,x_1,x_2)$ by
\begin{gather*}
T=t,\\
y_1 = \bfi\cdot[\bfC(s,t)+x_1\bfn(s,t)+x_2\bfb(s,t)],\\
y_2 = \bfj\cdot[\bfC(s,t)+x_1\bfn(s,t)+x_2\bfb(s,t)],\\
y_3 = \bfk\cdot[\bfC(s,t)+x_1\bfn(s,t)+x_2\bfb(s,t)].
\end{gather*}
The \wavetube apparatus follows integral curves of $\PD/\PD t$ in spacetime.
By expressing the vector field $\PD/\PD t$ with respect to
$(T,y_1,y_2,y_3)$:
\begin{equation*}
\frac{\PD}{\PD t} = \frac{\PD}{\PD T} +
\bfi\cdot\bfOm L_1 + \bfj\cdot\bfOm L_2 + \bfk\cdot\bfOm L_3
\end{equation*}
it can be seen that $\PD/\PD t$ is a linear combination of the
timelike future-pointing Killing vector field $K=\PD/\PD T$ and the
spacelike Killing vector field $L$:
\begin{equation*}
L = \bfi\cdot\bfOm L_1 + \bfj\cdot\bfOm L_2 + \bfk\cdot\bfOm L_3
\end{equation*}
where $[K,L]=0$.

%In terms of $(t,s,x_1,x_2)$ the metric tensor $g$ is
%\begin{gather}
%\notag
%g=-e^0\otimes e^0 + e^1\otimes e^1
%+e^2\otimes e^2 + e^3\otimes e^3 + O(\epsilon),\\
%\notag
%e^0 = \rho_0 d\Hlambda = c dt,\\
%\notag
%e^1 = \rho_0(d\Hs+\mu_2 d\Hlambda) = ds+\Omega a_0 dt,\\
%\notag
%e^2 = \rho_0 d\Hrho = d\rho,\\
%\label{lowest_order_Minkowski_coframe}
%e^3 = \rho_0\Hrho[d\phi + \mu_4 \Htau(\Hs) d\Hs] = \rho [d\phi +
%\tau(s)d s].
%\end{gather}
%where $\Set{e^0,e^1,e^2,e^3}$ is orthonormal to lowest
%order in $\epsilon$. Note that the approximate Minkowski metric is
%insensitive to $\mu_3$ and the curvature of $\bfC$.
At the instant
$t=t_0$ write $\bfC$ in the form
\begin{equation*}
\bfC(s,t_0) = \bfGamma_{t_0}(s) + \epsilon\bfgamma(s,t_0)
\end{equation*}
where $\bfGamma_{t_0}$ is a plane circle in $\BR^3$ rotating with
angular speed $\Omega$ about its axis $\bfk$.
The curve image $\bfGamma_{t_0}(s)$ has tangent $\bfT_{t_0}(s)$, normal
$\bfN_{t_0}(s)$, constant binormal $\bfB_{t_0}(s)=\bfk$ and radius $a_0$. Note
that $s$ is \emph{not} necessarily the arc parameter of
$\bfGamma_{t_0}$.

The vector
$\bfGamma_{t_0}(s)$ can be expressed
\begin{equation*}
\bfGamma_{t_0}(s) = -a_0\bfN_{t_0}(s)
\end{equation*}
and it is supposed that
the tangent $\bft$, normal $\bfn$ and binormal $\bfb$ of $\bfC$ satisfy
\begin{gather}
\notag
\bft(s,t) = \bfT_t(s) + O(\epsilon),\\
\notag
\bfn(s,t) = \cos[\beta(s,t)]\bfN_t(s) + \sin[\beta(s,t)]\bfB_t(s) +
O(\epsilon),\\
\label{moving_curve_requirements}
\bfb(s,t) = -\sin[\beta(s,t)]\bfN_t(s) + \cos[\beta(s,t)]\bfB_t(s) + O(\epsilon)
\end{gather}
for some angle $\beta(s,t)$. Using
\begin{gather*}
\bfT\cdot\bfk = 0,\\
\bfB = \bfk
\end{gather*}
it follows that
\begin{gather*}
\bft\cdot(\bfk\cross\bfC)(s,t) = a_0 + O(\epsilon),\\
\bfn\cdot(\bfk\cross\bfC)(s,t) = O(\epsilon),\\
\bfb\cdot(\bfk\cross\bfC)(s,t) = O(\epsilon)
\end{gather*}
which is consistent with (\ref{approximate_curve_conditions}).
%and so, by considering (\ref{approximate_curve_conditions}) it is seen that
%\begin{gather}
%\notag
%e^0 = c dt,\\
%\notag
%e^1 = ds + a_0\Omega dt,\\
%\notag
%e^2 = d\rho,\\
%\label{lowest_order_Minkowski_coframe_complete}
%e^3 = \rho[d\phi+\tau(s)ds].
%\end{gather}
Thus, in terms of $(t,s,x_1,x_2)$ the metric tensor $g$ is
\begin{gather}
\notag
g=-e^0\otimes e^0 + e^1\otimes e^1
+e^2\otimes e^2 + e^3\otimes e^3 + O(\epsilon),\\
\notag
e^0 = \rho_0 d\Hlambda = c dt,\\
\notag
e^1 = \rho_0(d\Hs+\mu_2 d\Hlambda) = ds+\Omega a_0 dt,\\
\notag
e^2 = \rho_0 d\Hrho = d\rho,\\
\label{lowest_order_Minkowski_coframe}
e^3 = \rho_0\Hrho[d\phi + \mu_4 \Htau(\Hs) d\Hs] = \rho [d\phi +
\tau(s)d s].
\end{gather}
where $\Set{e^0,e^1,e^2,e^3}$ is orthonormal to lowest
order in $\epsilon$. Note that the approximate Minkowski metric is
insensitive to $\mu_3$ and the curvature of $\bfC$.
\subsection{Slowly rotating Kerr spacetime}
%%%%%%%%%%%
The metric tensor at large distances from an isolated compact
rotating body in an asymptotically flat spacetime, with Newtonian
gravitational mass $M$ and angular momentum $J$, may be
approximated (for $r\gg 2\NewtonG M/c^2$ where $\NewtonG$ is
Newton's gravitational constant) by
\begin{equation}
\begin{split}
g =& -c^2\biggl(1-\frac{2\NewtonG M}{rc^2}\biggr)dT\Tensor dT\\
&+ \biggl(1+\frac{2\NewtonG M}{rc^2}\biggr)\bigl(dr\Tensor dr +
r^2d\theta\Tensor d\theta + r^2\sin^2\theta d\varphi\Tensor d\varphi\bigr)\\
&-\frac{2\NewtonG J}{r c^2}\sin^2\theta (d\varphi\Tensor dT +
dT\Tensor d\varphi)
\end{split}\label{kerrm}
\end{equation}
in spherical polar coordinates $(T,r,\theta,\varphi)$.

The use of the metric (\ref{kerrm}) is of course an idealisation
for physical applications such as terrestrial ring lasers, where
one does not have two Killing vectors. In particular, due to
interactions with other celestial bodies the positions of both
the rotation and angular momentum axes will not quite coincide
with each other and their positions will vary with time.
As reported in ref.\ \cite{LTR} for
the case of ring lasers at the Earth's surface
such effects are three to four orders of magnitude larger than
the Lense-Thirring effect. Nonetheless the calculation
presented here serves a useful purpose in isolating effects due to
gravito-magnetism. The implications of time variation of the rotation and
angular momentum axes of the Earth on terrestrial Lense-Thirring measurements
will be discussed elsewhere \cite{DLW}.

The vector fields $\PD/\PD T$ and $\PD/\PD\varphi$ are a commuting
pair of Killing vectors. Furthermore, $K=\PD/\PD T$ is timelike
and future-pointing and $L=\Omega\,\PD/\PD\varphi$ is spacelike
and has closed orbits. Consider the acceleration of matter on one of
the integral curves of the 4-velocity field
$V=(K+L)/\sqrt{-g(K+L,K+L)}$ with respect to a stationary observer
belonging to the 4-velocity field $W=K/\sqrt{-g(K,K)}$.
At events where the integral curves of $V$ and $W$ coincide $V$
has spatial acceleration ${\cal A}=H_W(\nabla_V V)$ where
$H_W\equiv 1+W\otimes \Mdual{W}$ projects the 4-acceleration to the
instantaneous 3-space of $W$. In the non-relativistic weak
gravitational field limit one finds that the g-orthonormal
components of ${\cal A}$ yield a centripetal acceleration with
dominant magnitude $r_0 \Omega^2$ at an event with coordinates
$r=r_0,\theta=\pi/2$. Thus each observer within $W$ would
interpret such events within $V$ to have instantaneous
3-acceleration produced by rotation with angular speed $\Omega$.

The asymptotically Minkowski coordinate
chart $(T,y_1,y_2,y_3)$ is given in terms of
$(T,r,\theta,\varphi)$ as
\begin{gather*}
y_1 = r\sin\theta\cos\varphi,\\
y_2 = r\sin\theta\sin\varphi,\\
y_3 = r\cos\theta.
\end{gather*}

The dot $\bfx\cdot\bfy$ and cross $\bfx\cross\bfy$ products
of the pair of vectors $\{\bfx,\bfy\}$ are defined with respect to the
tensor $h$:
\begin{equation*}
h = dy_1\Tensor dy_1 + dy_1\Tensor dy_1 + dy_2\Tensor dy_2 +
dy_3\Tensor dy_3
\end{equation*}
and the $h$-orthonormal frame $(\bfi,\bfj,\bfk)$:
\begin{gather*}
\bfi = \frac{\PD}{\PD y_1},\\
\bfj = \frac{\PD}{\PD y_2},\\
\bfk = \frac{\PD}{\PD y_3}
\end{gather*}
and, as before, the vector $\bfr_T$,
\begin{equation*}
\bfr_{T} = y_1\bfi + y_2\bfj + y_3\bfk,
\end{equation*}
locates the point $(T,y_1,y_2,y_3)$.

The metric tensor $g$ has the form
\begin{equation*}
\begin{split}
g =& -c^2\biggl(1-\frac{2\NewtonG M}{rc^2}\biggr)dT\Tensor dT +
\biggl(1+\frac{2\NewtonG M}{rc^2}\biggr)d\bfr\cdot d\bfr\\
&- \frac{2\NewtonG J}{r^3 c^2}\biggl[\bfk\cdot(\bfr\cross d\bfr)\Tensor dT
+ dT\Tensor \bfk\cdot(\bfr\cross d\bfr)\biggr]
\end{split}
\end{equation*}
where $(T,y_1,y_2,y_3)$ and the spacetime coordinates
$(t,\bvs,\bvx_1,\bvx_2)$ satisfy
\begin{gather*}
T=t,\\
y_1 = \bfi\cdot[\bfC(\bvs,t)+\bvx_1\bfn(\bvs,t)+\bvx_2\bfb(\bvs,t)],\\
y_2 = \bfj\cdot[\bfC(\bvs,t)+\bvx_1\bfn(\bvs,t)+\bvx_2\bfb(\bvs,t)],\\
y_3 = \bfk\cdot[\bfC(\bvs,t)+\bvx_1\bfn(\bvs,t)+\bvx_2\bfb(\bvs,t)]
\end{gather*}
and the $h$-orthonormal frame $(\bft,\bfn,\bfb)$ solves the Frenet-Serret
equations
\begin{gather*}
\bft = \frac{\PD}{\PD\bvs}\bfC,\\
\frac{\PD}{\PD\bvs}\bft = \bvkappa\bfn,\\
\frac{\PD}{\PD\bvs}\bfn = -\bvkappa\bft + \bvtau\bfb,\\
\frac{\PD}{\PD\bvs}\bfb = -\bvkappa\bfn.
\end{gather*}
The functions $\{\bvs,\bvx_1,\bvx_2,\bvkappa,\bvtau\}$
are underlined as a reminder
that they are normalized with respect to the tensor
$h$ rather than the spacetime metric $g$.
%the induced metric on the leaves of the constant $T$
%local foliation of spacetime.

As in the previous section consider curves of the form
\begin{equation*}
\bfC(\bvs,t) = \bfGamma_t(\bvs) + \epsilon \bfgamma(\bvs,t)
\end{equation*}
subject to (\ref{moving_curve_requirements}), with $s$ replaced by
$\bvs$, where $\bfGamma_t(\bvs)$ is the rotating plane circle of
$h$-radius $a_0$ centred at the point $\bfr_T=z_0\bfk,\quad z_0
>0$ with the Frenet frame $\{\bfT,\bfN,\bfB\}$ and
$\bfOm=\Omega\bfk$. Here one is considering $O(\epsilon)$
perturbations about a ``circular'' space curve parallel to the
``equatorial plane'' of the Kerr geometry where elements of the
\wavetube apparatus follow integral curves of $\PD/\PD t = \PD/\PD
T + \Omega\PD/\PD\varphi$ in this spacetime.

As before, introduce the dimensionless functions
\begin{gather*}
\bvHs = \frac{\bvs}{\rho_0},\\
\bvHx_1 = \frac{\bvx_1}{\rho_0},\\
\bvHx_2 = \frac{\bvx_2}{\rho_0},\\
\bvHkappa(\bvHs)=\frac{\bvkappa(\bvs)}{\kappa_0},\\
\bvHtau(\bvHs)=\frac{\bvtau(\bvs)}{\tau_0}.
\end{gather*}
Since
\begin{equation*}
\bfGamma_t(\bvs) = -a_0\bfN_t(\bvs)
\end{equation*}
and the position vector $\bfr$ has the form
\begin{equation*}
\bfr = -a_0 \bigl[\bfN + O(\epsilon)\bigr]
\end{equation*}
and using
\begin{gather*}\frac{d\bfr}{\rho_0} = \bft d\bvHs + \bfn [d\bvHx_1
-\mu_4\hat\bvtau(\bvHs) \bvHx_2 d\bvHs] + \bfb[d\bvHx_2 +\mu_4\hat\bvtau(\bvHs)
 \bvHx_1 d\bvHs] + \bft\mu_2 d\Hlambda,\\
\bfB = \bfk
\end{gather*}
with (\ref{moving_curve_requirements}) it follows that
\begin{equation*}
\bfk\cdot(\bfr\cross d\bfr) = \rho_0 a_0\bigl[d\bvHs +\mu_2 d\Hlambda
+ O(\epsilon)\bigr].
\end{equation*}

The metric tensor $g$ adapted to $\bfC$ is
\begin{equation}
\label{Kerr_metric}
\begin{split}
g/\rho_0^2 =& -(1-\mu_5)d\Hlambda\Tensor d\Hlambda + (1+\mu_5)\bigl[(d\bvHs +
\mu_2 d\Hlambda)\Tensor (d\bvHs + \mu_2 d\Hlambda)\\
&+
\bvHrho^2\bigl(d\phi+\mu_4\bvHtau(\bvHs)d\bvHs\bigr)\Tensor\bigl(d\phi+\mu_4\bvHtau(\bvHs)d\bvHs\bigr)
+ d\bvHrho\Tensor d\bvHrho\bigr]\\
&- \mu_6 \bigl[(d\bvHs + \mu_2 d\Hlambda)\Tensor d\Hlambda
+ d\Hlambda\Tensor (d\bvHs + \mu_2 d\Hlambda)\bigr]
\end{split}
\end{equation}
to lowest order in $\epsilon$ where
\begin{gather*}
r_0 = \sqrt{a_0^2+z_0^2},\\
\mu_5 = \frac{2\NewtonG M}{r_0 c^2},\\
\mu_6 = \frac{2\NewtonG J a_0}{r_0^3 c^3}
\end{gather*}
and
\begin{gather*}
\bvx_1 = \bvrho\cos\phi,\\
\bvx_2 = \bvrho\sin\phi.
\end{gather*}

A $g$-orthonormal coframe to lowest order in $\epsilon$ is
\begin{gather}
\notag
e^0 = \bigl[1-\mu_5+\mu_6^2/(1+\mu_5)]^{1/2}c
dt,\\
\notag
e^1 = \bigl[(1+\mu_5)^{1/2}\mu_2-\mu_6/(1+\mu_5)^{1/2}\bigr] cdt
+ ds,\\
\notag
e^2 = d\rho,\\
\label{lowest_order_Kerr_coframe}
e^3 = \rho [d\phi + \tau(s)ds]
\end{gather}
where
\begin{gather}
\notag
s = (1+\mu_5)^{1/2}\bvs,\\
\notag
\rho = (1+\mu_5)^{1/2}\bvrho,\\
\label{scales_and_barred_scales}
\tau(s) = (1+\mu_5)^{-1/2}\bvtau(\bvs).
\end{gather}
\section{Twisting EM modes on spacetime}
%%%%%%%%%%%
Both coframes (\ref{lowest_order_Minkowski_coframe}) and
(\ref{lowest_order_Kerr_coframe}) satisfy
(\ref{coframe_assumptions}) and so single-valued propagating
electromagnetic $F$ modes inside a \wavetube with circular cross-section of
radius $\rho_0$ can be generated from particular solutions of the
Helmholtz equations discussed earlier. Solutions regular
in the \wavetube follow from the forms
$\BesselJ{n}(|\alpha|\rho)\exp[i n\chi(\phi,s)]$ where $n\in\BZ$,
\begin{equation*}
\chi(s,\phi) = \phi + \int\limits^s_0\tau(\sigma)d\sigma
\end{equation*}
and $\alpha=\alpha_{TM}$ or $\alpha=\alpha_{TE}$ (see equations
(\ref{alpha_T_beta_T_explicit})).
The role of the Frenet torsion in solutions of the scalar
Helmholtz equation in coordinates adapted to non-planar curves
has been noted before in a number of different contexts
\cite{berry, kugler, inertial_guidance}.

The $F_{TM}$ and $F_{TE}$ modes follow from
\begin{equation*}
\Phi(s,\rho,\phi) = \BesselJ{n}(|\alpha_{TM}|\rho)\exp[i n\chi(\phi,s)]
\end{equation*}
with $\alpha_{TM}$ satisfying
\begin{equation*}
\BesselJ{n}(|\alpha_{TM}|\rho_0) = 0
\end{equation*}
and
\begin{equation*}
\Psi(s,\rho,\phi) = \BesselJ{n}(|\alpha_{TE}|\rho)\exp[i n\chi(\phi,s)]
\end{equation*}
with $\alpha_{TE}$ satisfying
\begin{equation*}
\BesselJ{n}^\prime(|\alpha_{TE}|\rho_0) = 0
\end{equation*}
where $\BesselJ{n}(x)$ is the $n$th regular Bessel function of the
first kind and $\BesselJ{n}^\prime(x) = d\BesselJ{n}(x)/dx$.

For both TE and TM modes
\begin{equation}
\label{xi_in_terms_omega_k}
\xi = -\omega_\xi dt + k_\xi ds
\end{equation}
and, using $V=\PD_t/\sqrt{-g(\PD_t,\PD_t)}$,
\begin{equation*}
\begin{split}
\nu_\xi &= \frac{c}{2\pi}\xi(V)\\
&= \frac{c\,\omega_\xi}{2\pi\sqrt{-g(\PD_t,\PD_t)}}
\end{split}
\end{equation*}
is the frequency (Hz), measured by the \wavetube apparatus, of the field mode
associated with $\xi=dW_{TM}$ or $\xi=dW_{TE}$.
Positive wavenumbers $k_\xi$ correspond to propagation in the
direction of the tangent $\bft$ and negative wavenumbers to
propagation in the opposite direction.

Since the \wavetube is closed, i.e. the \wavetube interior on spacetime is
topologically $\BR\cross S^1\cross \BD$, the Maxwell $2$-form satisfies
\begin{equation*}
F(t,s,\rho,\phi) = F(t,s+l,\rho,\phi),
\end{equation*}
where $l$ is the length of the \wavetube and so using
(\ref{xi_in_terms_omega_k}) and either
(\ref{TM_F_medium_in_terms_of_Phi}) or (\ref{TE_F_medium_in_terms_of_Psi})
\begin{equation}
\label{k_in_terms_of_N_tau}
k[N,n] \equiv k_\xi = \frac{2\pi N}{l} - n\Avtau,\quad\,N\in\BZ
\end{equation}
where
\begin{equation*}
\Avtau \equiv \frac{1}{l}\int\limits^l_0\tau(s)ds
\end{equation*}
is the average Frenet torsion. The mode spectrum is
classified using the triple of integers $(N,p,n)$, where the positive
integer $p$ labels a solution of $J_n(|\alpha_{TM}|\rho_0)=0$ in the TM case or
$J_n^\prime(|\alpha_{TE}|\rho_0)=0$ in the TE case.
\subsection{Frequency spectra}
%%%%%%%%%%%%
\label{section:frequency_spectra}
The longitudinal members of (\ref{lowest_order_Minkowski_coframe}) and
(\ref{lowest_order_Kerr_coframe}) have the form
\begin{gather*}
e^0 = A\, dt,\\
e^1 = B\, dt + ds
\end{gather*}
where $A$ and $B$ are constants with the dimensions of $c$.
For each $|\alpha|=\zeta/\rho_0$, where
$\alpha=\alpha_{TM}$ or
$\alpha=\alpha_{TE}$ (see equations (\ref{alpha_T_beta_T_explicit}))
and $\zeta$ is determined by the boundary
conditions, the dispersion relation $g^{-1}(\alpha,\alpha)=-|\alpha|^2$
with $\omega_\xi>0$ yields
\begin{equation}
\label{wavetube_omega_spectrum}
\omega_\xi =
\frac{\sqrt{A^2-B^2}}{A^2\Refind^2-B^2}
\biggl[-B\Kk+\sqrt{B^2\Kk^2
+(A^2\Refind^2-B^2)(\Kk^2+A^2\zeta^2/\rho^2_0)}
\biggr]
\end{equation}
where
\begin{equation*}
\Kk = k\sqrt{A^2-B^2}.
\end{equation*}
It is assumed that, for terrestrial applications, $|B|/A<\Refind$ and so the
sign of the square root of the discriminant in
(\ref{wavetube_omega_spectrum}) is chosen to be positive.
In addition to $k\equiv k[N,n]$ introduce the scalars $\{\omega_T,\zeta_T\}$
dependent on the mode indices $(N,p,n)$ and the type $T\in\Set{TE,TM}$ as follows:
\begin{gather*}
\omega_\xi \equiv \omega_T = \omega_T[N,p,n],\\
\zeta_T \equiv \zeta_T[p,n]
\end{gather*}
where $J_n(\zeta_{TM}[p,n])=0$ and
$J_n^\prime(\zeta_{TE}[p,n])=0$. The \wavetube apparatus measures the
frequency
\begin{equation*}
\nu_T[N,p,n] = \frac{c}{2\pi}dW_T\bigl(V)
\end{equation*}
and so here
\begin{equation}
\label{wavetube_frequency_spectrum}
\nu_T[N,p,n]=\frac{\omega_T[N,p,n]c}{2\pi\sqrt{A^2-B^2}}.
\end{equation}
We may approximate this expression further by taking into account the
relative magnitudes of $N$, $n$, $l\Avtau$, $\zeta_T$ and $k\rho_0$.
For a lasing medium excited by a RF field $|N|\gg 1$ is
expected and it is assumed that $2\pi|N|/l\gg|n\Avtau|$ and
$2\pi|N|/l\gg|\zeta_T|/\rho_0$. Then
(\ref{wavetube_omega_spectrum}) has the form
\begin{equation}
\label{high_freq_wavetube_omega_spectrum}
\omega_T[N,p,n] \simeq \frac{|k[N,n]|(A^2-B^2)}{A\Refind+\Sgn(k[N,n])B}
\end{equation}
where
\begin{equation*}
\Sgn(k) =
\begin{cases}
-1 & \text{if $k<0$,}\\
+1 & \text{if $k\ge 0$.}
\end{cases}
\end{equation*}
Using (\ref{k_in_terms_of_N_tau}) and $2\pi|N|/l >
|n\Avtau|$ it follows that $\Sgn(k)=\Sgn(N)$ hence
\begin{equation}
\label{mod_k_in_terms_of_N_tau}
|k|=\frac{2\pi|N|}{l}-\Sgn(N)n\Avtau.
\end{equation}
Note that $\omega_T$ and $\nu_T$ are independent of the mode type $T$ and the
Bessel root index $p$ in this approximation. Henceforth we write
$\omega=\omega_T$ and $\nu=\nu_T$ and drop their $p$ dependencies.

For terrestrial applications we expect that $|B|/A\ll 1$ (as well as
$|B|/A <\Refind$). In this case (\ref{wavetube_frequency_spectrum}) becomes
\begin{equation}
\label{approx_high_frequency_wavetube_frequency_spectrum}
\nu[N,n] \simeq \frac{c|N|}{l\Refind} -
\Sgn(N)\biggl[\frac{c|N|}{l}\frac{B}{A\Refind^2} +
\frac{cn\Avtau}{2\pi\Refind}\biggr]
\end{equation}
where (\ref{mod_k_in_terms_of_N_tau}) and
(\ref{high_freq_wavetube_omega_spectrum}) have been used.
\section{Multiple mode excitations and beating}
%%%%%%%%
If a number of electromagnetic modes of similar frequency are
active in the \wavetube then the time histories of the
electromagnetic fields measured by the \wavetube apparatus will exhibit
beating. The classical \cite{anderson_et_al:1994,stedman:1997}
Sagnac beat frequency $\delta\nu_\text{Sagnac}$
induced by rotation on flat spacetime follows from
(\ref{approx_high_frequency_wavetube_frequency_spectrum}) with $\Refind\simeq 1$ and
by choosing a pair of
rotationally symmetric ($n=0$) modes that differ only
in their directions of propagation and have $N>0$:
\begin{equation}
\label{explicit_Sagnac_beat_frequency}
\begin{split}
\delta\nu_\text{Sagnac} &\equiv \nu[N,0] - \nu[-N,0]\\
&\simeq -\frac{2\Omega N a_0}{l}
\end{split}
\end{equation}
where $A$ and $B$ have been obtained from
(\ref{lowest_order_Minkowski_coframe}).
To reveal the classical
expression for the Sagnac beat frequency the parameters $a_0$ and $l$
and the mode index $N$ are eliminated by introducing the area $\Ar\equiv \pi a_0^2$
and perimeter $\Pe\equiv 2\pi a_0$ of the plane circle approximated by the wavetube locus
and the lasing wavelength $\lambda\equiv\l/N$. It follows that
\begin{equation}
\label{standard_Sagnac_beat_frequency}
\delta\nu_\text{Sagnac} \simeq -\frac{4\Omega\Ar}{\lambda\Pe}.
\end{equation}
\subsection{Twisted mode beating on flat spacetime}
%%%%%%%%%%%
The most general frequency difference is
constructed from a pair of modes of type $T_1,T_2\in\Set{TM,TE}$ with
indices $(N_1,p_1,n_1)$ and $(N_2,p_2,n_2)$. In the approximation
introduced earlier we only need consider $N$ and $n$ dependences:
\begin{equation*}
\delta\nu[N_1,n_1;N_2,n_2] \equiv
\nu[N_1,n_1] - \nu[N_2,n_2].
\end{equation*}
Thus, with $N>0$, (\ref{standard_Sagnac_beat_frequency}) has the generalization
\begin{equation}
\label{generalized_Sagnac_beat_frequency}
\delta\nu[N,n;-N,n] \simeq
-\frac{2N\Omega a_0}{l\Refind^2} - \frac{cn\Avtau}{\pi\Refind}.
\end{equation}

For terrestrial applications the external RF field generates counter-
and co-propagating modes
with indices $N_1$
and $N_2$ whose magnitudes are so large it is possible that
$|N_1|\neq|N_2|$. More generally, there is no reason to assume that
$(|N_1|,p_1,n_1,T_1)$ and $(|N_2|,p_2,n_2,T_2)$ are identical.
With $N_1>0$ and $N_2<0$ and using
(\ref{approx_high_frequency_wavetube_frequency_spectrum}) it follows that
\begin{equation*}
\begin{split}
\delta\nu[N_1,n_1;N_2,n_2] \simeq&
\frac{c}{l\Refind}(N_1+N_2) - \frac{a_0\Omega}{l\Refind^2}(N_1-N_2)\\
&- \frac{c\Avtau}{2\pi\Refind}(n_1+n_2).
\end{split}
\end{equation*}
\subsection{Twisted mode beating on slowly rotating Kerr spacetime}
%%%%%%%%%%%
The dimensionless parameters $\{\mu_2,\mu_6\}$,
\begin{gather*}
\mu_2 = \frac{a_0\Omega}{c},\\
%\mu_5 = \frac{2\NewtonG M}{r_0 c^2},\\
\mu_6 = \frac{2\NewtonG J a_0}{r_0^3 c^3},
\end{gather*}
contain the \emph{Euclidean} radius $a_0$ of the multiply-wound plane circle
$\bfGamma$ that approximates the wavetube locus on Kerr spacetime
$(\Mcal{M},g)$. One may eliminate the parameter $a_0$ in favour of a
length determined by the spacetime metric $g$ rather than the
Euclidean metric $d\bfr\cdot d\bfr$.

For each constant $(T_0,r_0,\theta_0)$ the multiply-wound planar
circle may be described as the image of the map
\begin{equation}
\label{spacetime_Gamma_map}
\begin{split}
\bfGamma : [0,2\pi m] &\rightarrow \Mcal{M}\\
u &\mapsto (T=T_0,\,r=r_0,\,\theta=\theta_0,\,\varphi=u)
\end{split}
\end{equation}
where $(T,r,\theta,\varphi)$ are the coordinates used in (\ref{kerrm}),
$a_0=r_0\sin\theta_0$ and the positive integer $m$ is a winding
number. The arclength $l_0$ of $\bfGamma$ is
\begin{equation}
\label{g_length_of_circle}
l_0 \equiv \int\limits^{2\pi m}_0
\sqrt{g(\bfGamma^\prime,\bfGamma^\prime)} du
\end{equation}
where $\bfGamma^\prime = \PD_\varphi$ is tangent to
$\bfGamma$. Using (\ref{kerrm}), (\ref{spacetime_Gamma_map}) and
(\ref{g_length_of_circle}) it follows that
\begin{equation}
\label{a0_in_terms_of_l0}
a_0 = \frac{l_0}{2\pi m} \frac{1}{\sqrt{1+\mu_5}}
\end{equation}
where
\begin{equation*}
\mu_5 = \frac{2\NewtonG M}{r_0 c^2}.
\end{equation*}
Introduce dimensionless parameters $\{\mu_2^\prime,\mu_6^\prime\}$
\begin{gather*}
\mu_2^\prime \equiv \frac{\Omega}{c}\frac{l_0}{2\pi m},\\
\mu_6^\prime \equiv \frac{2\NewtonG J}{r_0^3 c^3}\frac{l_0}{2\pi m}
\end{gather*}
so that, by using (\ref{a0_in_terms_of_l0}) to eliminate $a_0$:
\begin{gather*}
\mu_2 = \frac{\mu_2^\prime}{\sqrt{1+\mu_5}},\\
\mu_6 = \frac{\mu_6^\prime}{\sqrt{1+\mu_5}}.
\end{gather*}

Typically, for terrestrial applications, the magnitudes of the dimensionless parameters
$\{\mu_2,\mu_5^\prime,\mu_6^\prime\}$ are much less than unity. Thus, it is
reasonable to approximate $B/A$ by a polynomial obtained from the
multivariate Taylor expansion of $B/A$ with respect
$(\mu_2,\mu_5^\prime,\mu_6^\prime)$ about $(0,0,0)$.
However, when truncating the Taylor
expansion the numerical values of the terms in the series must be carefully
scrutinized. For example, the parameters associated with a wavetube
whose locus approximates $\bfGamma$ with ``radius'' $l_0/(2\pi m) \simeq
150\w{m}$ fixed on the Earth
%and centred on the poles have magnitudes $|\mu_2^\prime|\simeq 10^{-10}$,
%$|\mu_5|\simeq 10^{-8}$ and $|\mu_6^\prime|\simeq 10^{-19}$. For terrestrial
and centred on the poles have magnitudes
$|\mu_2^\prime|\simeq 3.6\times10^{-11}$,
$|\mu_5|\simeq 1.4\times10^{-9}$ and
$|\mu_6^\prime|\simeq 1.7\times10^{-20}$. For terrestrial
conditions $|\mu_2^\prime\mu_5|>|\mu_6^\prime|$ is generally true. It follows that
\begin{equation}
\label{approx_Kerr_B_over_A}
\frac{B}{A} \simeq \mu_2^\prime + \frac{1}{2}\mu_2^\prime\mu_5 - \mu_6^\prime
\end{equation}
is a good approximation (the other terms in the Taylor
series are about 10 orders of magnitude smaller).

Using (\ref{approx_high_frequency_wavetube_frequency_spectrum}) and
(\ref{approx_Kerr_B_over_A}) it follows that
\begin{equation}
\label{approx_Kerr_high_frequency_spectrum_explicit}
\nu[N,n]
\simeq \frac{c|N|}{l\Refind}
-\Sgn(N)\biggl[\frac{l_0}{2\pi m}\frac{\Omega|N|}{l\Refind^2}
\biggl(1+\frac{\NewtonG M}{r_0 c^2}
-\frac{2\NewtonG J}{\Omega r_0^3 c^2}
\biggl) + \frac{cn\Avtau}{2\pi\Refind}\biggr]
\end{equation}
and, more generally, for a pair of modes with indices $(N_1,n_1)$ and $(N_2,n_2)$,
where $N_1>0$ and $N_2<0$,
\begin{equation}
\label{generalized_Kerr_Sagnac_beat_frequency}
\begin{split}
\delta\nu[N_1,n_1;N_2,n_2] \simeq&
\frac{c}{l\Refind}(N_1+N_2)\\ &-\frac{l_0}{2\pi m}\frac{\Omega}{l\Refind^2}
\left[1+{\NewtonG M\over r_0 c^2}-{2\NewtonG J\over\Omega r_0^3 c^2}\right]
(N_1-N_2)\\
&- \frac{c\Avtau}{2\pi\Refind}(n_1+n_2).
\end{split}
\end{equation}

%It can be seen (\ref{approx_Kerr_refractive_index_spectrum_explicit}) has the
%same structure as the frequency spectrum on flat spacetime except that
%$\Omega$ is replaced by $\Omega+\NewtonG M\Omega/(r_0 c^2)-2G_NJ/(r_0^3c^2)$.
%In fact, since $\Omega={d\varphi\over dt}$ is the angular velocity measured
%at spatial infinity, it differs from the angular velocity measured at $r_0$ by
%a gravitational time dilation factor. In terms of the angular velocity
%$\Omega_0$ measured at $r_0$, which would correspond, for example, to the
%standard angular velocity of the Earth measured at its surface, the
%corresponding expression is simply $\Omega_0-2G_NJ/(r_0^3c^2)$ to
%leading order.
%Using (\ref{approx_high_frequency_wavetube_frequency_spectrum})
%the frequency difference of a pair of modes with indices $(N_1,n_1)$
%and $(N_2,n_2)$, where $N_1>0$ and $N_2<0$, is
%\begin{equation*}
%\begin{split}
%\delta\nu[N_1,n_1;N_2,n_2] \simeq&
%\frac{c}{l\Refind}(N_1+N_2)\\ &-\frac{2\Ar\Omega}{l\Pe\Refind^2}
%\left[1+{\NewtonG M\over r_0 c^2}-{2G_NJ\over\Omega r_0^3c^2}\right]
%(N_1-N_2)\\
%&- \frac{c\Avtau}{2\pi\Refind}(n_1+n_2)
%\end{split}
%\end{equation*}

By choosing a pair of
rotationally symmetric ($n=0$) modes that differ only
in their directions of propagation and have $N_1=-N_2=N>0$ we find the
classical Sagnac beat frequency (\ref{standard_Sagnac_beat_frequency})
is modified in this approximation to
\begin{equation}
\begin{split}
\delta\nu_\text{Sagnac}
&= \nu[N,0] - \nu[-N,0]\\
&\simeq -\frac{1}{\Refind^2}\frac{4\Omega\Ar}{\lambda\Pe}
\left[1+{\NewtonG M\over r_0 c^2}-{2G_NJ\over\Omega r_0^3c^2}\right]
\end{split}
\end{equation}
where the ``area'' $\Ar$ and ``perimeter'' $\Pe$ of $\bfGamma$ are
defined as
\begin{gather*}
\Ar \equiv \frac{l_0^2}{4\pi m^2},\\
\Pe \equiv \frac{l_0}{m}
\end{gather*}
in terms of total arclength $l_0$ and winding number $m$ of $\bfGamma$
and $\lambda=l_0/N$.
%At the
%Earth's surface the magnitude of the ``gravitational time dilation''
%correction is $\NewtonG M/(r_0 c^2)= 7.0\times10^{-10}$, while the
%gravito-magnetic correction is
%$-2G_NJ/(\Omega r_0^3c^2)=-4.6\times10^{-10}$. The next largest
%corrections are at the level of $10^{-12}$ of the base Sagnac frequency,
%arising from the ``special relativistic time dilation''.

Using (\ref{scales_and_barred_scales}) it can be seen that the average torsion
$\Avtau$ is related to the Euclidean torsion $\bvtau(\bvs)$ of the
wavetube locus by
\begin{equation*}
\Avtau = \frac{1}{\sqrt{1+\mu_5}}\frac{1}{\bvl}\int\limits^\bvl_0
\bvtau(\bvs)d\bvs
\end{equation*}
where $\bvl \equiv l/\sqrt{1+\mu_5}$.
\subsection{Example}
%%%%%%%%%%%
\begin{figure}
\begin{center}
\scalebox{0.75}{\includegraphics{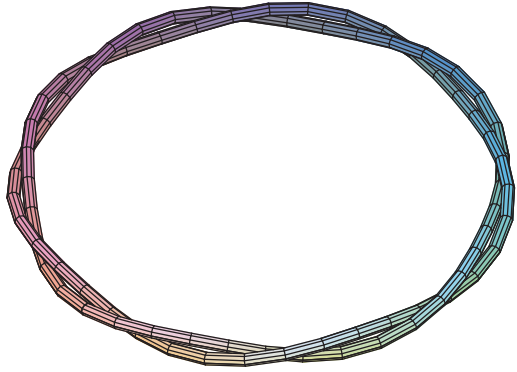}}
\caption{\label{figure:knot}A 3-dimensional diagram of the type
$(2,9;150,5,5)$ torus knot.}
\end{center}
\end{figure}
To illustrate the significance of the Frenet torsion contribution
to (\ref{generalized_Sagnac_beat_frequency}) we consider a
\wavetube based on a particular torus knot. \Subfignum
\begin{figure}
\begin{center}
\scalebox{0.4}{\includegraphics{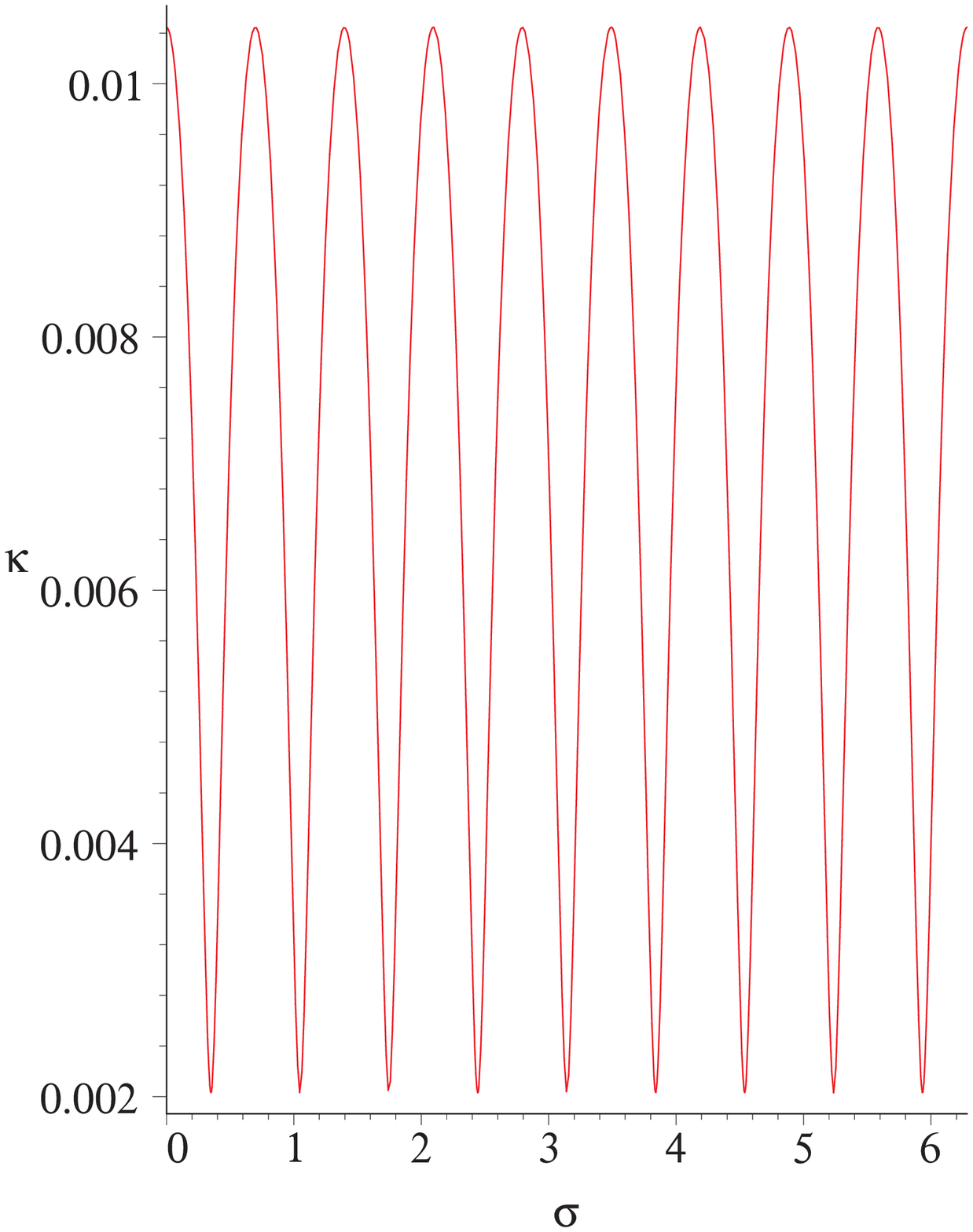}}
\caption{\label{figure:curvature}The Euclidean Frenet curvature
$\bvkappa$ of the type $(2,9;150,5,5)$ torus knot. Note that the
abscissa is labeled by $\sigma$ and not the Euclidean arclength
$\bvs$ (see main text).} \Stepfigc
\scalebox{0.4}{\includegraphics{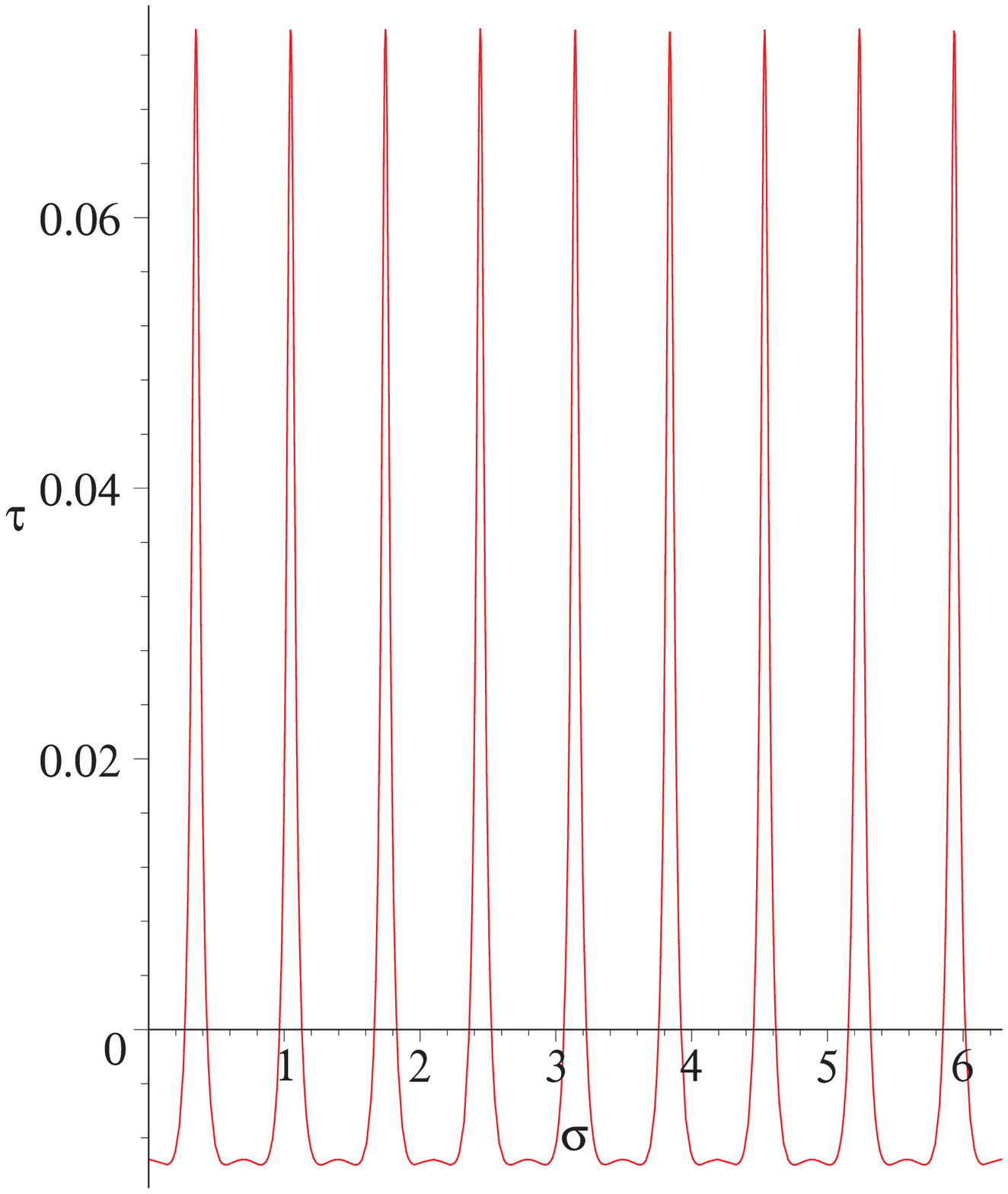}}
\caption{\label{figure:torsion}The Euclidean Frenet torsion
$\bvtau$ of the type $(2,9;150,5,5)$ torus knot. Note that the
abscissa is labeled by $\sigma$ and not the Euclidean arclength
$\bvs$ (see main text).}
\end{center}
\end{figure}
\Normfignum
A type $(u_1,u_2;v_1,v_2,v_3)$ torus knot is a curve $\TorusCarc$
specified by a pair of integers $(u_1,u_2)\in\BZ^2$ and a triple of real numbers
$(v_1,v_2,v_3)\in\BR^3$ :
\begin{gather*}
\TorusC : \sigma \to \bigl[\TorusC_1(\sigma),
\TorusC_2(\sigma),\TorusC_3(\sigma)]\\
\TorusC_1(\sigma) = \bfi\cdot\TorusC(\sigma) = [v_1 +
v_2\cos(u_2\sigma)]\cos(u_1\sigma),\\
\TorusC_2(\sigma) = \bfj\cdot\TorusC(\sigma) = [v_1 +
v_2\cos(u_2\sigma)]\sin(u_1\sigma),\\
\TorusC_3(\sigma) = \bfk\cdot\TorusC(\sigma) = v_3\sin(u_2\sigma),\\
\TorusCarc(\bvs) = \TorusC\bigl(\sigma(\bvs)\bigr)
\end{gather*}
where the Euclidean arc parameter $\bvs$ satisfies
\begin{equation*}
\frac{d\TorusCarc}{d\bvs}\cdot\frac{d\TorusCarc}{d\bvs}=1.
\end{equation*}
The torus knot $\TorusCarc$ is approximated by a multiply-wound circle of
Euclidean radius $a_0=v_1$.

The type $(2,9;150,5,5)$ torus knot is shown in figure
\ref{figure:knot} and its Frenet curvature and torsion are shown
in figures \ref{figure:curvature} and \ref{figure:torsion}. Choosing MKS units,
so $a_0=v_1=150\w{m}$, this particular knot has
$\Avtau=3.13\times10^{-4}\w{m}^{-1}$ and so $cn\Avtau/\pi = n 29.8\w{kHz}$.
The classical Sagnac beat frequency (\ref{standard_Sagnac_beat_frequency})
of a circular He--Ne wavetube laser ($\lambda = 633\w{nm}$) and ``radius''
$150\w{m}$ fixed on the Earth ($\Omega=7.29\times10^{-5} \w{rad}/\w{s}$)
and centred on the poles is $34.5\w{kHz}$. Clearly, for
$n=1$ and $\Refind\simeq 1$ the standard Sagnac and torsion
contributions to the beat frequency are quite similar and, moreover,
in order to determine $\Omega$ from the beat frequency both
contributions would have to be taken into account.

%To have strong confidence in the lowest $\epsilon$-order results derived
%above the constants $\Set{\mu_1,\mu_2,\mu_3,\mu_4}$ must
%satisfy $\mu_3,\mu_1\ll\mu_2,\mu_4$. Since $\mu_2=3.9\times 10^{-9}$ it
%follows that $\rho_0\ll\kappa_0\mu_2=3.5\times 10^{-9}m$ which,
%clearly, is far too small to be practical. The situation becomes even
%more dramatic when considering the values of $\Set{\mu_5,\mu_6,\mu_7}$.
%However, the approximation
%scheme \emph{can} be used consistently by assigning an
%$\epsilon$-order higher that $1$ to $\mu_2$.
%The $F$ modes derived above can then be adopted as the leading
%$\epsilon$-order terms in a perturbation hierachy that extends to the
%desired $\epsilon$ order. This approach,
%although well-posed and quite simple in principle, is certainly
%onerous in practice. However, it is certain that the Frenet torsion will still
%significantly contribute to the beat frequency and we expect that the
%above example is at least qualitatively correct for a physically
%reasonable value of $\rho_0$.
\section*{Conclusions}
%%%%%%%%
A particular approximation scheme has been developed that permits
one to determine the electromagnetic mode structure for a rotating
slender wavetube containing a non-conducting isotropic homogeneous dispersionless medium.
The possible effects of a weak non-Newtonian gravito-magnetic
field on these modes has been included. The approximation enables
one to identify TE and TM type field configurations with respect
to the longitudinal axis of the wavetube. Although this need not
be fixed in space, provided the dimensionless parameters $\mu_1$
and $\mu_3$ are smaller than $\mu_2$ the decomposition remains
valid to the order prescribed. In this regime one finds that such
a wavetube with non-zero integrated Frenet torsion can produce a
modification to the classical Sagnac beat frequency due to any
rotation of the interferometer.

To detect effects of gravito-magnetism in this scenario one must
adjust the parameters $\Set{l_0,\Omega,\rho_0,\tau_0,\kappa_0}$ so
that the effects due to the dimensionless parameters
$\Set{\mu_5,\mu_6^\prime}$ can be distinguished experimentally from
those produced by $\Set{\mu_1,\mu_2^\prime,\mu_3}$. There are of
course practical limitations that bound $l_0$ and $\rho_0$ and the
possibility of increasing $\Omega$ by rotating the interferometer.
Furthermore the example given in section 7.1 shows that there
exist slender geometrical structures based on space curves having
small Frenet curvature but integrated Frenet torsion giving rise
to interference effects commensurate with those produced by the
rotation of the Earth. It is difficult to construct geometrical
configurations that meet all the above competing constraints
needed to reveal terrestrial gravito-magnetic effects without
releasing the condition of relatively constant local Frenet
curvature.

To overcome this limitation it is clear that a more complete
analysis of the mode spectra should take into account
perturbations induced by variations of Frenet curvature of the
wavetube. This is a non-trivial modification since it implies that
the Maxwell equations no longer separate into twisted TE and TM
modes. However, for appropriate local curvature variations one may
extend the approximation scheme using a perturbation approach and
the basis of twisted modes found in this analysis. This would
yield a clear separation of the effects of Frenet torsion and
Frenet curvature on the Sagnac beat frequency. Such developments
would provide a more complete picture of the delicate interplay
between acceleration, geometry, electromagnetism and gravity that
arises in any attempt to design an effective ring-laser capable of
detecting terrestrial gravito-magnetism.

\section*{Acknowledgements}
%%%%%%%%%
DAB, AN and RWT are most grateful for the hospitality provided by the
Department of Physics and Astronomy, University of Canterbury, Christchurch,
New Zealand and for
valuable discussions with G Stedman, R Hurst and R Reeves. They are also
grateful to M Hamilton, University of Adelaide, for information on optical
polarising devices. DAB acknowledges financial support from the
Royal Society, DLW from the Marsden fund of the Royal Society of New
Zealand and RWT is grateful to BAe Systems for their support
and interest in this research.

%---------------------------------------------

\end{document}